\documentclass{article}

\usepackage{PRIMEarxiv}

\usepackage[utf8]{inputenc} 
\usepackage[T1]{fontenc}    
\usepackage{hyperref}       
\usepackage{url}            
\usepackage{booktabs}       
\usepackage{amsfonts}       
\usepackage{nicefrac}       
\usepackage{microtype}      
\usepackage{lipsum}
\usepackage{wrapfig}
\usepackage{natbib}
\usepackage{fancyhdr}       
\usepackage{graphicx}       
\graphicspath{{media/}}     

\title{Uncertainty Propagation in LLM-Based Systems}

\author{
Boming Xia$^{1,2,3}$,
Liming Zhu$^{3,4}$,
Erdun Gao$^{1,2,3}$,
Qinghua Lu$^{3,4}$,
Minhui Xue$^{3,1}$,
Dino Sejdinovic$^{1,2}$ \\[0.5em]
$^{1}$Responsible AI Research (RAIR) Centre \\
$^{2}$Adelaide University \\
$^{3}$CSIRO \\
$^{4}$UNSW Sydney \\
Australia
}

\begin{document}
\maketitle

\begin{abstract}
Uncertainty in large language model (LLM)-based systems is often studied at the level of a single model output, yet deployed LLM applications are compound systems in which uncertainty is transformed and reused across model internals, workflow stages, component boundaries, persistent state, and human or organisational processes. Without principled treatment of how uncertainty is carried and reused across these boundaries, early errors can propagate and compound in ways that are difficult to detect and govern. This paper develops a systems-level account of uncertainty propagation. It introduces a conceptual framing for characterising propagated uncertainty signals, presents a structured taxonomy spanning intra-model (P1), system-level (P2), and socio-technical (P3) propagation mechanisms, synthesises cross-cutting engineering insights, and identifies five open research challenges.
\end{abstract}


\section{Introduction}
\label{sec:intro}

Large language models (LLMs) are increasingly deployed not as isolated predictors but as components in larger software systems that integrate model inference with external resources, executable tools, verification steps, and persistent state~\cite{qiu2024llm,chen2025standalone,bass2025engineering}. These range from focused integration patterns such as retrieval-augmented generation (RAG)~\cite{lewis2020retrieval}, tool-using pipelines~\cite{li2025review}, and verifier--generator cascades such as LLM-as-a-judge~\cite{zheng2023judging,zhugeagent}, to skill-augmented agents~\cite{jiang2026sok} and fully orchestrated multi-step workflows that plan, act, observe outcomes, and update state across extended runs~\cite{yao2022react,shinn2023reflexion}. The rapid practical deployment of such systems across enterprise, scientific, and consumer settings has outpaced the development of principled frameworks for reasoning about their failure modes. A distinctive threat in such compound LLM-based systems (LLM systems hereinafter) is \emph{cascading failure}: intermediate artefacts produced under weak evidence can be committed to state, routed into tools, or used to justify later decisions, allowing early errors to compound through the run~\cite{zhu2025llm,zhou2025shielda,zhang2025agentracer}.

Cascading failure is difficult to analyse because compound systems repeatedly convert uncertain intermediate evidence into later actions, artefacts, and control decisions. At different points in a run, systems may produce or expose token-level probabilities, verifier scores, retrieval coverage indicators, or user-facing textual hedges. We refer to all such measurable or recorded quantities that are treated as evidence of uncertainty by a downstream consumer as \emph{uncertainty signals}. The scope of this survey is defined not by signal type but by the \emph{propagation role} a signal plays: what occurs when it crosses a boundary to a new consumer, form, or decision context. Such crossings are not neutral transmissions. The scope of a signal may narrow or widen, its encoding may change in ways that alter precision or coverage, its consumer may hold different calibration assumptions, and its decision role may shift from diagnostics to runtime control. LLMs compound this difficulty by producing fluent outputs even when prompts are underspecified or outside the model's effective knowledge, and a substantial body of work documents the resulting hallucination, miscalibration, and overconfidence~\cite{xiongcan,kalai2024calibrated,kalai2025language,chhikara2025mind}. The engineering question is therefore not only whether an uncertainty signal can be produced at a given step, but what happens as it moves from the context in which it was estimated to the components and actors that must act on it.

To help address this gap, this survey focuses on \emph{uncertainty propagation} in LLM systems.  We define propagation as the transmission of uncertainty signals across boundaries at which their scope, form, consumer, or decision role changes. The definition is contrastive: \emph{estimation} concerns how a signal is produced; \emph{calibration} concerns correcting its magnitude against empirical outcomes; \emph{communication} concerns rendering it interpretable to a human recipient. Propagation concerns what occurs across those steps: what a signal is treated as being about at each point, where it is made available for downstream use, how it is used at each decision point, and whose uncertainty it is taken to express and who acts on it. This survey draws on work in estimation, calibration, and communication, but the organising question is the cross-boundary trajectory of the signal rather than any single step within it.

We organise the literature by where an uncertainty signal is consumed after it is produced, distinguishing three levels: \textbf{P1 (intra-model, within-request)}, where a signal is produced and consumed within a single model-facing request; \textbf{P2 (system-level, cross-component)}, where a signal is consumed by a technical component of the deployed system to shape subsequent execution; and \textbf{P3 (socio-technical, beyond-system)}, where a signal is consumed by a user, organisation, or auditor outside the deployed technical system.\footnote{We use \emph{deployed technical system} to denote the software components and workflow-integrated services whose outputs are automatically consumed as part of system operation, treating end users, organisational processes, and real-world outcomes as beyond this boundary. By a \emph{single model-facing request} we mean one request issued at the model-facing boundary of the surrounding system, including any internal generation, scoring, or sampling steps that remain internal to that request and are resolved before any non-model component acts on the associated uncertainty signal.} Consumption locus determines what a signal must be: coherence within a generation step, interface-expressibility at P2, and attributable interpretability at P3. It determines where failures are recoverable: a signal lost at a P2 interface cannot be restored by P3 governance. And it reveals cross-level failure patterns that signal-type or task-domain partitions of the literature structurally cannot see. Section~\ref{sec:conceptual} formalises these boundaries and specifies the placement rules used throughout.

\begin{figure}
    \centering
    \includegraphics[width=0.75\linewidth]{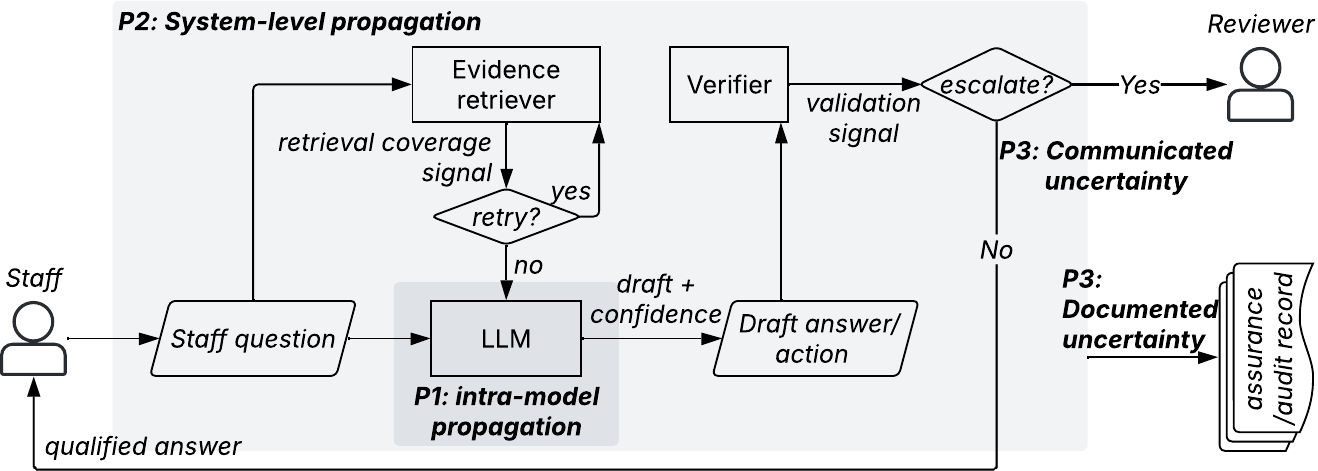}
    \caption{Illustrative example of uncertainty propagation in an LLM-based policy and compliance assistant.}
    \label{fig:example}
\end{figure}

Consider an LLM-based policy and compliance assistant that answers staff questions using retrieval together with verification (Figure~\ref{fig:example}). A staff member asks whether a proposed vendor contract term is permissible under current procurement policy. The system retrieves relevant policy clauses; if retrieval coverage is insufficient, that signal triggers a retry before generation proceeds (\textbf{P2}). Otherwise the system prompts the LLM with the retrieved clauses to draft a response, whose generation carries an internal confidence signal before any non-model component acts on it (\textbf{P1}). The draft is then passed to a verifier, which produces a validation signal the system uses to decide whether to escalate to a human reviewer or return the response directly (\textbf{P2}). If not escalated, the system returns a qualified response and appends a record to the audit log documenting the confidence level under which the answer was produced; if escalated, the response is held for the reviewer (\textbf{P3}). The point is not the particular compliance workflow, but the propagation structure it exposes: the same evidential state crosses model, component, human, and organisational boundaries, and at each boundary its form, consumer, and decision role may change. The same structure is not specific to compliance assistants; it recurs across compound LLM systems whenever generated artefacts, tool results, tests, verification records, or human review decisions become inputs to later action. This survey organises these mechanisms into a taxonomy of uncertainty propagation across P1, P2, and P3.

The survey makes three main contributions:
\begin{itemize}
  \item \textbf{A conceptual framework for uncertainty propagation (\S\ref{sec:conceptual}).} We introduce a vocabulary for characterising uncertainty signals at boundaries: the object of concern, the decision role, the reuse locus, and the attribution-consumer pair. The framework supplies the placement rules that organise the taxonomy and enables precise description of cross-boundary failure modes that existing survey vocabularies cannot express.

  \item \textbf{A structured taxonomy of propagation patterns spanning P1--P3 (\S\ref{sec:taxonomy}).} For each pattern, we characterise the signal types involved, the boundary crossed, the tuple fields that change, and the downstream consequences for how the signal can be interpreted and acted on by its next consumer.

  \item \textbf{Cross-cutting engineering principles and a prioritised research agenda (\S\ref{sec:insights} and \S\ref{sec:gaps}).} We identify recurring cross-cutting insights and derive engineering implications for LLM system design, evaluation, and governance. We also structure five open research challenges to inform future research.
\end{itemize}

\subsection*{Positioning relative to adjacent surveys} Adjacent surveys provide three important foundations for this review. First, surveys on LLM uncertainty estimation and quantification organise the space by how uncertainty signals are obtained, including verbalised confidence, model-internal cues such as token probabilities and latent probes, sampling disagreement, perturbation-based measures, and calibration or selective prediction protocols~\citep{xia2025survey,shorinwa2025survey,kang2025uncertainty,huang2024survey,he2025survey,liu2025uncertainty,abbasli2025comparing}. Second, surveys on RAG, tool use, agentic workflows, and multi-agent protocols organise compound LLM systems by architecture, workflow structure, memory, tool interfaces, orchestration, and interaction patterns~\citep{gao2024ragllm,xu2025llm,wang_survey_2024,guo2024multiagent,fan2024survey,huang2024understanding,xi2025rise}. Third, reliability and hallucination surveys consolidate detection and mitigation pipelines, including grounding, verification, critique, revision, and judge-based mechanisms~\citep{ye2023cognitive,zhang2025siren,huang2025hallucination,pan2025towards,tonmoy2024survey}. These bodies of work establish the signal repertoire, system substrate, and mitigation practices on which uncertainty propagation depends. However, none treats propagation itself as its primary organising question: how uncertainty signals are carried forward, re-expressed, retained, and consumed across model, system, and socio-technical boundaries. Our survey therefore complements adjacent work by taking uncertainty propagation as the organising focus, asking not only how uncertainty is produced or used locally, but how it is reused across downstream decision points.

\subsection*{Organisation}
Section~\ref{sec:conceptual} introduces the conceptual framing, defining uncertainty propagation contrastively relative to estimation, calibration, and communication, and specifying the vocabulary used throughout the taxonomy. Section~\ref{sec:taxonomy} presents the P1--P3 taxonomy, characterising how uncertainty signals are produced, transformed, and consumed across intra-model, system-level, and socio-technical boundaries. Section~\ref{sec:insights} synthesises cross-cutting insights including recurring semantic failure modes, the expressiveness-compatibility tension between richer signals and interface constraints, and the asymmetric development of the literature across P1, P2, and P3, and derives engineering implications for system design, evaluation, and governance. Section~\ref{sec:gaps} identifies five open research challenges spanning semantic fidelity, propagation-quality evaluation, uncertainty-to-action semantics, socio-technical uptake, and formal compositional foundations. Section~\ref{sec:conclusion} summarises the main claims and their implications for engineering LLM systems under uncertainty.

\section{Conceptual Framing for Uncertainty Propagation}
\label{sec:conceptual}

This section introduces the vocabulary used throughout the survey to analyse uncertainty propagation. The introduction defined an \emph{uncertainty signal} as any measurable or recorded quantity treated as evidence of uncertainty by a downstream consumer. A signal alone, however, does not determine meaning: the same signal may be treated as uncertainty about different targets, used for different downstream actions, and made explicit at different points in a system or organisation. Propagation analysis therefore requires tracking not just signals but \emph{contextualised uncertainty statements}: a signal together with the information needed to interpret and act on it. We formalise this as a tuple \(u = \langle s, o, d, r, (a,c) \rangle\), where \(s\) is the uncertainty signal, \(o\) is the object of concern, \(d\) is the decision role, \(r\) is the reuse locus at which the statement is made explicit, \(a\) is the attribution whose uncertainty the statement is taken to express, and \(c\) is the actor that consumes the statement at a decision point. This is a descriptive abstraction for comparing heterogeneous systems, not a full epistemic model. The following subsections define each field and then illustrate the notation on a concrete trace.

\subsection{Uncertainty signals and their scope}
\label{sec:concept_signal}

The field \(s\) in the tuple denotes the \textbf{uncertainty signal} available for downstream use at a given point in execution. Representative signals include token-level probabilities or entropy, disagreement across samples, retrieval or tool diagnostics, and verifier or judge scores. The signal need not be a scalar confidence score: some approaches use richer forms such as probability intervals, credal sets, or possibility functions, which preserve distinctions that a point estimate would collapse~\cite{yang2026verbalizing}. Whether scalar, interval, label, disagreement pattern, or structured diagnostic, \(s\) denotes whatever is actually available for downstream use at the point in question.

This matters because signals are often produced at one scope and consumed at another. A component may emit a token-level, span-level, sample-level, or step-level signal, while the decisions that consume it are typically taken at the level of a response, workflow step, run, or operating condition~\citep{he2025survey}. Signals are therefore often aggregated, thresholded, or re-expressed before they can be consumed downstream. The remaining fields of the tuple, \(o\), \(d\), \(r\), and \((a,c)\), record the context that turns a signal into a reusable uncertainty statement: what it is taken to be about, how it is used, where it is made explicit, and by whom.

\subsection{Object of concern}
\label{sec:concept_object}

The \textbf{object of concern} \(o\) identifies what the uncertainty statement is treated as being about at a given point in execution. The categories below captures recurrent target classes in the surveyed literature, not an exhaustive ontology.

\begin{itemize}
    \item \textbf{Output artefact (\(o_{\mathrm{art}}\)).} The statement concerns an artefact produced or assembled by the system, such as an answer, a retrieved evidence set, a plan, a tool result, or a verifier assessment. The uncertainty is typically about whether that artefact is adequate, reliable, or sufficiently supported for its intended use.

    \item \textbf{Decision input or prerequisite (\(o_{\mathrm{input}}\)).} The statement concerns whether the information needed to proceed is sufficiently specified. Examples include underspecified intent, missing constraints, missing facts, applicability conditions, or incomplete acceptance criteria.

    \item \textbf{Operating conditions and controls
    (\(o_{\mathrm{op}}\)).} The statement concerns whether the
    available evidence, safeguards, oversight arrangements, or
    operating constraints are sufficient to support deployment or continued operation under stated conditions.
\end{itemize}

During propagation, the object of concern can shift without the signal itself changing, for instance from an output artefact to operating conditions as run-level evidence accumulates across a workflow. Making \(o\) explicit is what allows such shifts to be tracked rather than left implicit.

\subsection{Decision role}
\label{sec:concept_decision}

The \textbf{decision role} \(d\) identifies how the uncertainty statement is used to guide action at a given decision point. It does not encode a full objective function or control policy; it captures the mode of use at a given decision point. Decision roles attach to a use event, not intrinsically to the signal itself. We distinguish five main recurrent roles across the surveyed literature:

\begin{itemize}
    \item \textbf{Diagnosis (\(d_{\mathrm{diag}}\)).} The statement is used to inspect, rank, triage, or warn about likely error or risk without by itself directly determining the next action.

    \item \textbf{Runtime control (\(d_{\mathrm{ctrl}}\)).} The statement is used to directly determine what happens next within the current execution, for example whether to accept, retrieve, verify, retry, revise, abstain, defer, or escalate.

    \item \textbf{Cross-run adaptation (\(d_{\mathrm{adapt}}\)).} The statement is used to update behaviour across runs or over time, for example to recalibrate thresholds, rebalance sampling, or revise routing and review policies.

    \item \textbf{Communication (\(d_{\mathrm{comm}}\)).} The statement is used to communicate uncertainty in a form intended to shape reliance, checking, oversight, or coordination by another actor.

    \item \textbf{Assurance support (\(d_{\mathrm{assure}}\)).} The statement is used as evidence in governance or operational decisions, such as release or authorisation decisions, operating restrictions, rollback or mitigation decisions, or comparative claims with stated scope and coverage.
\end{itemize}

Different decision roles require different properties from the uncertainty statement. For example, diagnosis relies mainly on useful ordering or triage value; runtime control requires threshold stability or reliable abstention behaviour; communication requires faithful and interpretable presentation; and assurance support depends more on scope, traceability, and coverage than on numerical calibration alone. The same signal may therefore be adequate for one role and inadequate for another. Critically, the decision role can change independently of the object of concern during propagation: a statement may remain about an output artefact while its role shifts from diagnosis to runtime control and later to communication. Making \(d\) explicit allows such shifts to be identified and evaluated rather than left implicit.

\subsection{Reuse locus}
\label{sec:concept_reuse}

The \textbf{reuse locus} \(r\) identifies where an uncertainty statement is made explicit in a form that technical or human actors can act on. This is distinct from who uses it: \(r\) records where the statement becomes available, while the consumer field \(c\) records who acts on it at a decision point. A statement may become available at one locus and be consumed later by a different actor or at a different stage of the workflow. We distinguish four recurrent loci:

\begin{itemize}
    \item \textbf{Model-internal (\(r_{\mathrm{int}}\)).} The statement is produced and consumed within a single model-facing request, before any non-model component acts on it. Typical examples include token-level probabilities, internal confidence scores, or disagreement signals that remain within one request.

    \item \textbf{System-held (\(r_{\mathrm{sys}}\)).} The statement is explicit in machine-readable technical state that later components can access and reuse. Examples include response-level aggregates, verifier outputs, retrieval or tool diagnostics, flags, thresholds, and structured annotations attached to workflow state.

    \item \textbf{Human-facing (\(r_{\mathrm{hum}}\)).} The statement is re-expressed for human interpretation, for example as a warning, confidence cue, ranked alternative, explanation, or visual summary. At this locus, the statement can shape reliance, checking, intervention, or escalation.

    \item \textbf{Assurance artefacts (\(r_{\mathrm{assure}}\)).} The statement is recorded in artefacts that persist beyond a single run, such as monitoring summaries, audit trails, incident records, evaluation reports, or risk register entries. At this locus, the statement can support review, accountability, and longer-horizon operational decisions.
\end{itemize}

The reuse locus directly grounds the P1--P3 distinction. P1 concerns statements that remain model-internal, produced and consumed before any non-model component acts on them. P2 concerns statements externalised into system-held state and consumed by downstream technical components to steer execution. P3 concerns statements that reach human-facing or assurance-artefact loci, where they are taken up by users, operators, or organisational processes. These locus definitions supply the placement rules used throughout the taxonomy: a mechanism is classified at P1, P2, or P3 according to the highest locus its uncertainty statements reach. A statement well-formed at \(r_{\mathrm{int}}\) but never externalised to \(r_{\mathrm{sys}}\) cannot influence P2 execution; a statement available at \(r_{\mathrm{sys}}\) but never surfaced to \(r_{\mathrm{hum}}\) or \(r_{\mathrm{assure}}\) cannot support P3 governance.

\subsection{Attribution and consumption}
\label{sec:a2}

The \textbf{attribution} \(a\) identifies the actor whose uncertainty the statement is taken to express. The \textbf{consumer} \(c\) identifies the actor that uses the statement at a decision point. In this survey, the relevant actors are the model, the surrounding system, a human, or an assurance actor (such as an auditor, operator, or governance body). These are recurrent actor roles in the surveyed literature, not an exhaustive ontology.

Attribution need not coincide with the component that computed the signal, and need not coincide with the consumer. A response-level score derived from model behaviour may be attributed to the model even when consumed by the surrounding system. A statement may be attributed to the system when it summarises retrieval quality, tool reliability, or accumulated evidence across components, and later consumed by a human or assurance actor as it crosses a stakeholder or governance boundary. Making \(a\) and \(c\) explicit is what allows a change in whose uncertainty is being asserted to be distinguished from a change in who is acting on it, shifts that can and do occur independently during propagation.

\subsection{Illustrative Trace of Uncertainty Propagation}
\label{sec:concept_example}

We illustrate the tuple on the policy-and-compliance assistant from the introduction \S\ref{sec:intro}. A staff member asks: \emph{``Can I approve a laptop-purchase exception for a contractor under the travel budget policy?''} The system retrieves procurement and travel-policy passages, but coverage is insufficient: the retrieved set addresses employee thresholds but says little about contractor exceptions. The system records a coverage diagnostic in system-held state and uses it to decide whether to retry retrieval or proceed to generation. This is already a P2 signal: uncertainty produced by a system component and consumed by the system to steer execution. We focus the formal trace on the model confidence signal that arises during generation, which illustrates the full P1\(\to\)P2\(\to\)P3 arc most clearly.

\textbf{P1: Model-internal confidence.}
During drafting, the model produces an internal confidence signal
over its generated response:
\(
u_{\mathrm{int}} = \langle s_{\mathrm{conf}},\, o_{\mathrm{art}},\, d_{\mathrm{ctrl}},\, r_{\mathrm{int}},\, (\mathrm{model}, \mathrm{model}) \rangle.
\)
The signal is attributed to and consumed by the model (\(a = c = \mathrm{model}\)), concerns the draft as an output artefact (\(o_{\mathrm{art}}\)), and remains model-internal (\(r_{\mathrm{int}}\)) before any non-model component acts on it.

\textbf{P1 \texorpdfstring{$\to$}{->} P2: Externalisation.} The model-internal signal is re-expressed into a system-usable form:
\(
u_{\mathrm{ctrl}} = \langle s_{\mathrm{draft}},\, o_{\mathrm{art}},\, d_{\mathrm{ctrl}},
\, r_{\mathrm{sys}},\, (\mathrm{model}, \mathrm{sys}) \rangle,
\)
with \(u_{\mathrm{int}} \to u_{\mathrm{ctrl}}\). Two fields change: the reuse locus advances from model-internal to system-held (\(r_{\mathrm{int}} \to r_{\mathrm{sys}}\)), and the consumer shifts from the model to the system (\(c: \mathrm{model} \to \mathrm{sys}\)). The statement is now available to steer branching, abstention, or escalation.

\textbf{P3: Human-facing communication.}
If support remains weak, the system surfaces a warning such as \emph{``Policy coverage for contractor exceptions appears incomplete; human review recommended.''} This yields:
\(
u_{\mathrm{hum}} = \langle s_{\mathrm{warn}},\, o_{\mathrm{input}},\, d_{\mathrm{comm}},\, r_{\mathrm{hum}},\,
(\mathrm{sys}, \mathrm{human}) \rangle,
\)
with \(u_{\mathrm{ctrl}} \to u_{\mathrm{hum}}\). Three fields change simultaneously: the locus advances to human-facing (\(r_{\mathrm{hum}}\)), the object shifts from the answer artefact to whether the decision can proceed given missing prerequisites (\(o_{\mathrm{input}}\)), and the role shifts from runtime control to communication (\(d_{\mathrm{comm}}\)). What steered system execution now guides human judgement.

\textbf{P3: Assurance artefact.}
Repeated low-coverage cases of this kind may be aggregated into a governance record noting that contractor-related queries frequently require escalation. Such a record instantiates \(u_{\mathrm{assure}}\) with \(o_{\mathrm{op}}\), \(d_{\mathrm{assure}}\), \(r_{\mathrm{assure}}\), and \((\mathrm{sys}, \mathrm{assure})\): the object has shifted to operating conditions, the role to assurance support, and the consumer to an assurance actor. Uncertainty is no longer steering the present interaction; it is evidence for governance.

\textbf{What the trace shows.}
Each transition changes a subset of the tuple fields while leaving others stable. The reuse locus advances monotonically from model-internal through system-held to human-facing and assurance artefact. The object of concern and decision role shift at boundaries where the purpose of the signal changes. The attribution-consumer pair records whose uncertainty is being expressed and who acts on it at each stage.

\begin{figure}
    \centering
    \includegraphics[width=1\linewidth]{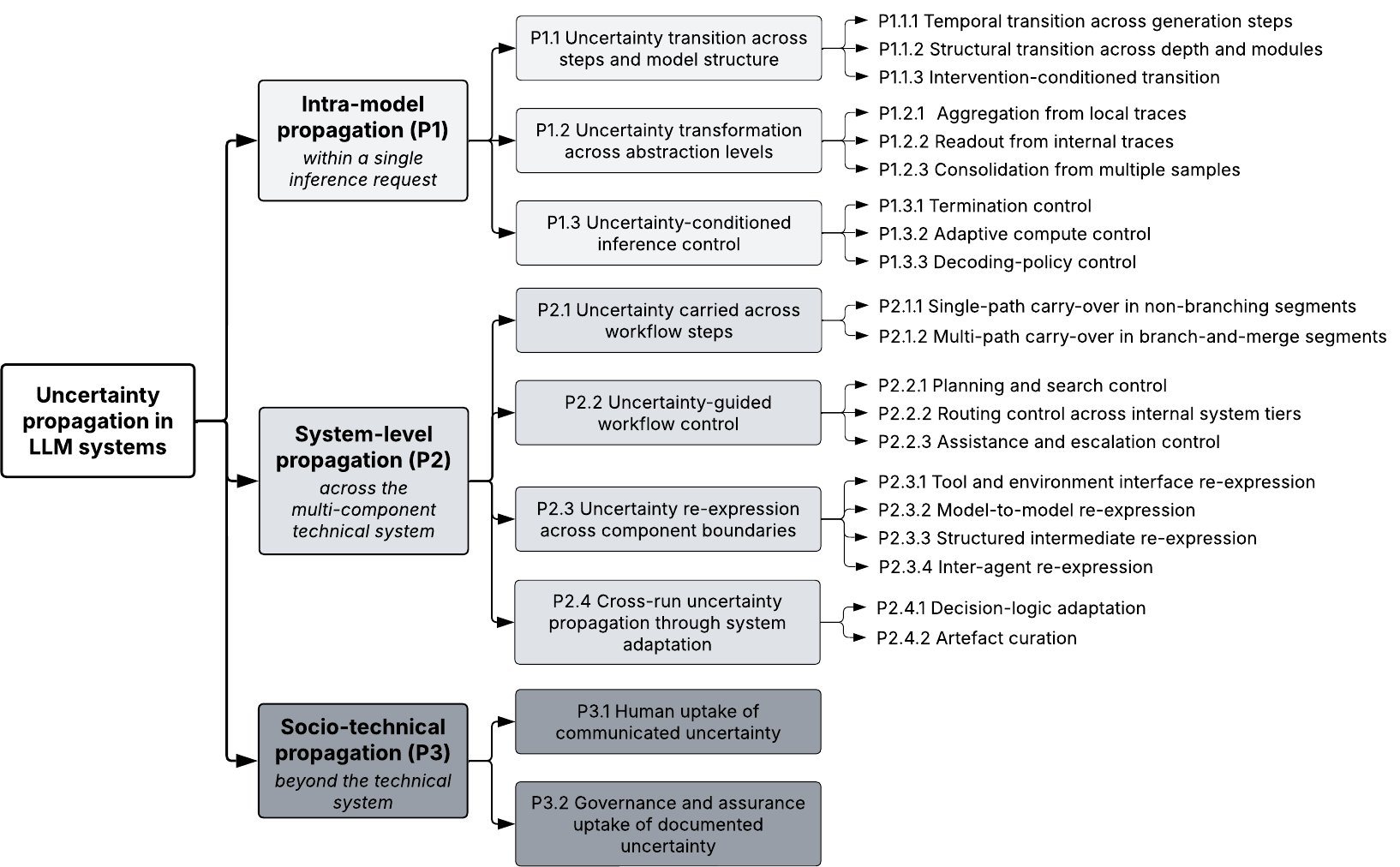}
    \caption{Overview of the taxonomy of uncertainty propagation in LLM systems.}
    \label{fig:taxo_overview}
\end{figure}

\section{Taxonomy of Uncertainty Propagation in LLM Systems}
\label{sec:taxonomy}

This section develops a taxonomy of recurring uncertainty propagation patterns in LLM systems. Building on Section~\ref{sec:conceptual}, we classify propagation mechanisms by three questions: which fields of the uncertainty statement \(u=\langle s,o,d,r,(a,c)\rangle\) change at a boundary, which boundary is crossed, and how the resulting statement is used downstream. Figure~\ref{fig:taxo_overview} provides a high-level overview. The taxonomy is organised first by boundary: \textbf{P1} covers propagation within a single model-facing request; \textbf{P2} covers propagation across components and execution steps within the deployed technical system; and \textbf{P3} covers propagation beyond the technical system, where uncertainty statements are taken up by users, operators, or organisational and governance processes. Within each level, we distinguish recurring propagation classes.

\subsection{Within-request uncertainty propagation (P1)}
\label{sec:intra-model}

P1 covers uncertainty propagation within a single model-facing
request. It includes internal samples, candidates, passes, or decoding trajectories so long as they remain internal to that request and are not externalised into reusable technical state or consumed by any non-model component. In the terminology of Section~\ref{sec:conceptual}, the uncertainty statement remains at the model-internal reuse locus \(r_{\mathrm{int}}\). Once an uncertainty signal is made explicit in system-held state that later components or steps can consume, the case falls under P2 rather than P1.

Within this boundary, the main propagation patterns concern internal carry-over, within-request re-expression, and inference-time control. Uncertainty may be propagated across generation steps, depth, or modules (\hyperref[sec:p11]{P1.1}); re-expressed into a different proxy or abstraction level within the same request (\hyperref[sec:p12]{P1.2}); or reused to shape decoding, compute allocation, or stopping behaviour (\hyperref[sec:p13]{P1.3}).

\begin{figure}
    \centering
    \includegraphics[width=0.8\linewidth]{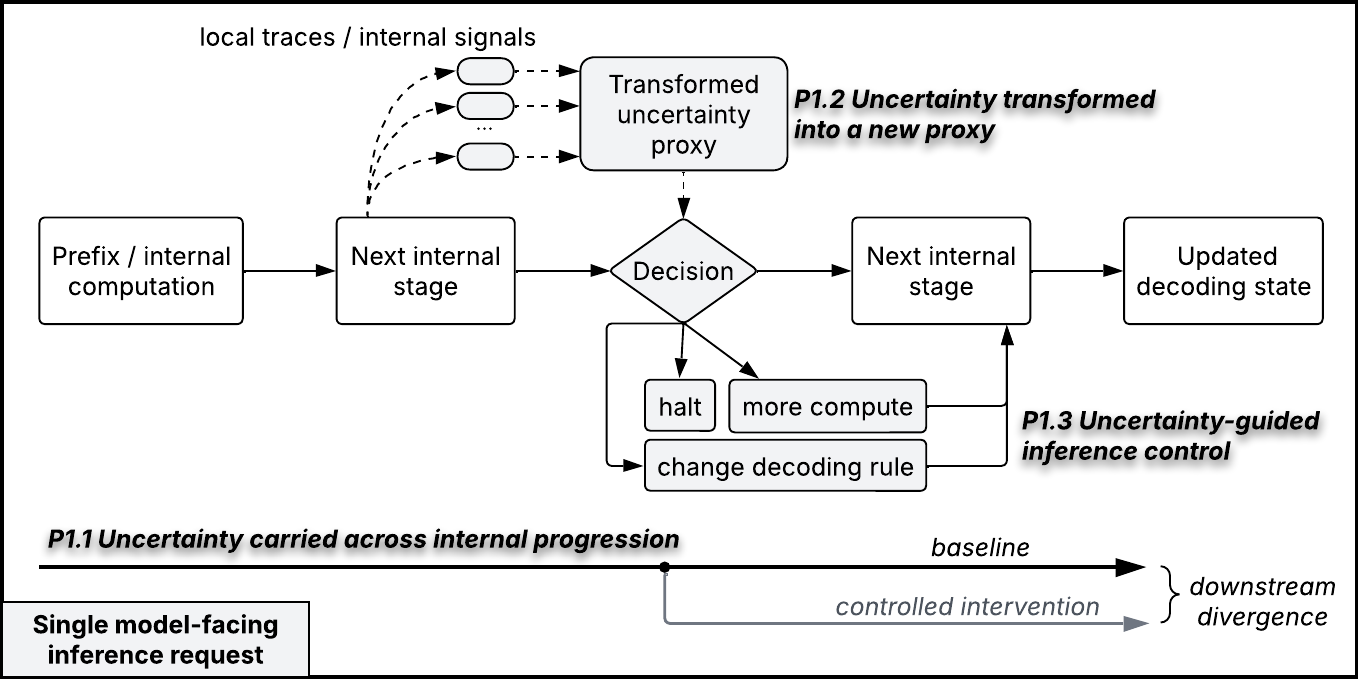}
    \caption{Schematic overview of intra-model uncertainty propagation in P1.}
    \label{fig:p1_overview}
\end{figure}

\subsubsection{Uncertainty transition across sequential internal positions (P1.1)}
\label{sec:p11}

\begin{wrapfigure}{r}{0.55\textwidth}
    \centering
    \includegraphics[width=\linewidth]{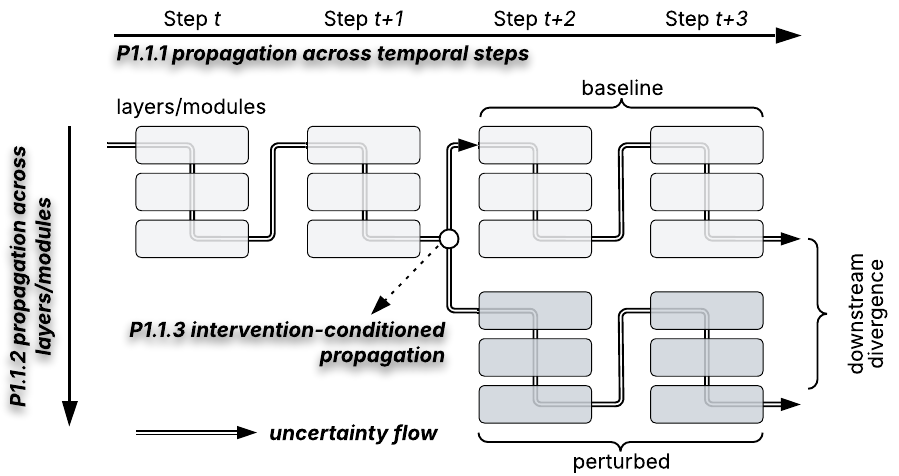}
    \caption{Uncertainty transition (P1.1)}
    \label{fig:p1.1}
\end{wrapfigure}

P1.1 covers within-request propagation in which an uncertainty signal is traced as it evolves along an ordered sequence of internal positions within a single model-facing request, such as generation steps, model depth, branches, or modules (see Figure \ref{fig:p1.1}). The statement remains at \(r_{\mathrm{int}}\) throughout. This distinguishes P1.1 from P1.2, where a signal is transformed into a new proxy of a different type within the same request, and from P1.3, where a signal is consumed to control how computation proceeds.

The canonical P1.1 case involves two successive positions within the same request, written \(u \to u'\) where \(u=\langle s,o,d,r,(a,c)\rangle\) and \(u'=\langle s',o,d,r,(a,c)\rangle\). The salient update is \(s \to s'\); the object of concern, reuse locus, and attribution--consumer pair are treated as fixed for the transition under analysis. The decision role is typically \(d_{\mathrm{diag}}\): the analysis is oriented toward understanding how the signal evolves rather than toward directly controlling inference. When a within-request signal is instead consumed to determine the next control action, the case is better placed in P1.3.

What propagates in P1.1 is an uncertainty signal as it accumulates, decays, or shifts character across an ordered internal sequence. Typical instances include step-indexed uncertainty dynamics during decoding (\hyperref[sec:p111]{P1.1.1}), stage-wise instability or error accumulation through depth or modules (\hyperref[sec:p112]{P1.1.2}), and intervention-conditioned divergence where a controlled perturbation reshapes later internal transitions (\hyperref[sec:p113]{P1.1.3}).

\paragraph{Temporal transition across generation steps (P1.1.1)}
\label{sec:p111}

P1.1.1 covers within-request propagation in which an uncertainty signal is traced across successive generation steps of a single model-facing request. The defining feature is that the sequential evolution of the signal across token positions is analytically material: the contribution must treat how the signal changes across positions as its object of study, rather than merely reporting uncertainty values at isolated positions.
For successive decoding steps \(t\) and \(t+1\), we write \(u_t \to u_{t+1}\), where the salient update is \(s_t \to s_{t+1}\) and the decision role is typically \(d_{\mathrm{diag}}\). The object of concern, reuse locus, and attribution--consumer pair are usually treated as fixed for the transition under analysis.

Two patterns recur most often in this subtype. One is \emph{path dependence}, a property of the estimation mechanism: the signal at step \(t+1\) depends on earlier signal values or decoding choices, so the estimate at each position cannot be interpreted as an isolated per-token quantity. The other is \emph{position-sensitive dynamics}, an empirical finding about the distribution of influence: some token positions have disproportionate downstream effect on the signal trajectory, so that uniform per-position treatment is misaligned with the underlying dynamics. The two patterns are distinct and can occur independently: a method can exhibit path dependence without position-sensitive dynamics, and position-sensitive dynamics can be identified without a path-dependent estimator.

The literature instantiates these temporal patterns in several ways. Some works construct token-level or stepwise uncertainty estimates with explicit dependence on prior steps, so that the estimate at step \(t+1\) is not determined solely by the instantaneous next-token distribution \citep{vazhentsev2025unconditional,vazhentsev2025uncertainty,lee2025uncertainty}. Other works identify that error and reliability dynamics are concentrated at a sparse set of high-impact positions rather than distributed uniformly across tokens, so that some positions function as points at which the distribution over plausible continuations changes abruptly~\citep{arbuzov2025beyond}.

A useful boundary case is a token-indexed trace that is later collapsed into a response-level proxy, for example \citet{moslonka2025learned}. Such work is best placed in P1.1.1 only when the main analytic focus is the temporal transition dynamics themselves, rather than the aggregation into a coarser proxy, which belongs in P1.2.1.

\paragraph{Structural transition across depth and modules (P1.1.2)}\label{sec:p112}

P1.1.2 covers within-request propagation in which an uncertainty signal is traced across ordered structural stages of a single model-facing request, such as transformer depth or module sequence. The defining feature is that stage-to-stage evolution is analytically material: the contribution must treat how the signal changes as it passes through successive layers or modules as its object of study, rather than using those stages only as alternative sites at which to extract a signal. For successive stages \(\ell\) and \(\ell+1\), the salient update is \(s_\ell \to s_{\ell+1}\) with decision role \(d_{\mathrm{diag}}\); the object of concern, reuse locus, and attribution--consumer pair are usually fixed.

The structural evolution in P1.1.2 tends to take two complementary forms. \emph{Depth-wise carry-over} refers to systematic change in the uncertainty signal as internal signals pass through successive layers or modules: each stage leaves a traceable imprint on what the next stage receives. \emph{Depth-wise dominance} refers to the uneven distribution of that influence: a subset of layers or modules disproportionately shapes the signal's trajectory, so that treating all stages as equally contributory misrepresents the underlying dynamics. Carry-over describes the propagation mechanism, dominance describes where in the structure that mechanism has the greatest effect.

Several distinct instantiations of these patterns appear in the literature. Some works treat layer or module depth as integral to the propagation mechanism rather than as a convenient probe location~\citep{tan2025bottom}, , while others incorporate cross-layer or layer-specific uncertainty signals into a broader uncertainty pipeline~\citep{huang2025reppl,vazhentsev2025uncertainty}. A specialised instance arises at the level of arithmetic precision: \citep{budzinskiy2025numerical} analyses round-off error propagation in a decoder-only transformer by composing block-level error bounds across depth, instantiating the depth-wise carry-over pattern in the domain of numerical fidelity rather than semantic uncertainty.

A useful boundary case is a method that records layer-indexed signals but whose main contribution is reading out a proxy from those signals rather than analysing the stage-to-stage carry-over itself; such work belongs in readout transformation (P1.2.2). If the main contribution instead lies in pooling layer-indexed signals into a coarser proxy, it belongs in aggregation transformation (P1.2.1).

\paragraph{Intervention-conditioned transition (P1.1.3)}
\label{sec:p113}

P1.1.3 covers within-request propagation in which a controlled intervention alters the subsequent trajectory of a model-facing request, and the resulting downstream divergence serves as the uncertainty signal. The defining feature is intervention-conditioned divergence: a baseline realisation and one or more deliberately perturbed variants are compared under the same request context, and uncertainty is operationalised as the divergence between them.

Let \(u_i^{\mathrm{base}}\) and \(u_i^{\mathrm{pert}}\) denote the uncertainty statements for the baseline and perturbed variants at a later comparison position. The salient signal is a divergence measure between \(s_i^{\mathrm{base}}\) and \(s_i^{\mathrm{pert}}\), serving as the signal field \(s\) of the resulting uncertainty statement. The decision role is typically \(d_{\mathrm{diag}}\), since intervention-conditioned divergence analysis is primarily oriented toward understanding where and how intervention-induced divergence arises rather than directly controlling inference.
The object of concern, reuse locus, and attribution--consumer pair are treated as fixed for the transition
under analysis.

Two forms of intervention-conditioned divergence recur across the literature. \emph{Continuation sensitivity} arises when a targeted intervention at a decoding prefix or internal state reshapes the set of plausible later continuations, so that the divergence between baseline and perturbed variants captures how much the trajectory depends on that prefix or state. \emph{Source attribution} arises when one source of information or internal component is perturbed while others are held fixed, attributing the resulting downstream divergence to that specific source or interaction.

Both are reflected in the literature. For continuation sensitivity, controlled interventions on intermediate generative states have been shown to induce downstream divergence in later outcome distributions, with the induced divergence treated as the primary signal rather than as an incidental evaluation artefact~\citep{zur2025language}. For source attribution, controlled variation of one input source while holding others fixed can isolate source-specific uncertainty contributions, while joint variation captures interaction-level contribution across sources~\citep{tang2025analysis}. Across both patterns, the perturbation target determines what the induced divergence is taken to measure: trajectory sensitivity in the first case, source-specific or interaction contribution in the second.

A useful boundary case arises when multiple perturbed variants are collapsed into a single proxy: such work belongs in P1.1.3 when the controlled intervention and induced divergence are the main analytic object, and in multi-sample consolidation (P1.2.3) when the main contribution lies in consolidating multiple realisations into one proxy regardless of how they were generated. When the induced divergence is consumed to steer within-request computation rather than to characterise uncertainty diagnostically, the case belongs in P1.3.

\subsubsection{Uncertainty transformation across abstraction levels (P1.2)}
\label{sec:p12}

\begin{wrapfigure}{r}{0.4\textwidth}
    \centering
    \includegraphics[width=\linewidth]{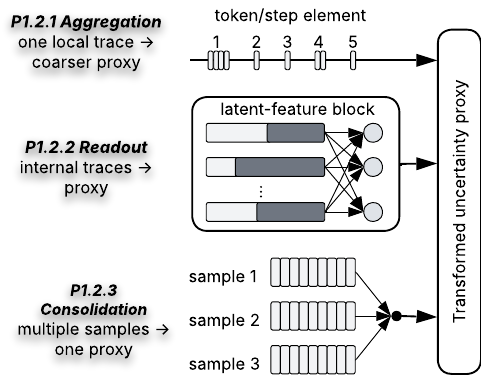}
    \caption{Uncertainty transformation (P1.2)}
    \label{fig:p1.2}
\end{wrapfigure}

P1.2 covers within-request propagation in which an uncertainty signal is transformed into a proxy of a different type or scope within a single model-facing request (see Figure \ref{fig:p1.2}). Where P1.1 traces how a signal evolves along an ordered sequence of internal positions, P1.2 concerns how a signal is re-expressed into a new form: aggregated from local traces, read out from internal features, or consolidated across parallel realisations. The statement remains at \(r_{\mathrm{int}}\) throughout.

For two within-request uncertainty statements, we write \(u \to u'\), where the salient update is \(s \to s'\): a within-request uncertainty signal is re-expressed into a new proxy, with the decision role typically \(d_{\mathrm{diag}}\) and the object of concern, reuse locus, and attribution--consumer pair treated as fixed for the transformation under analysis.

What propagates in P1.2 is uncertainty re-expressed into a new within-request proxy, changing form without leaving the model-internal locus: aggregated from local token- or step-indexed traces (\hyperref[sec:p121]{P1.2.1}), read out from model-internal states (\hyperref[sec:p122]{P1.2.2}), or consolidated across multiple within-request realisations
(\hyperref[sec:p123]{P1.2.3}).

\paragraph{Aggregation from local traces (P1.2.1)}
\label{sec:p121}

P1.2.1 covers within-request transformation in which local token- or step-indexed uncertainty signals are aggregated into a coarser proxy within a single model-facing request. The defining move is the aggregation itself: local evidence is combined according to a defined rule to produce a proxy at span, entity, or response level.
For a within-request trace of local statements \(\{u_i\}_{i=1}^{n}\), we write \(\{u_i\}_{i=1}^{n} \to u_{\mathrm{agg}}\), where the salient transformation is \(\{s_i\}_{i=1}^{n} \to s_{\mathrm{agg}}\). The decision role of the resulting statement is typically \(d_{\mathrm{diag}}\): the aggregated proxy is produced for diagnostic use within the request.

Two aspects of the aggregation recur as analytically distinct. \emph{Combination rule} refers to how local signals are combined across positions: whether by sequential updating, weighted summation, or other schemes that assign differential weight to positions. \emph{Aggregation scope} refers to which signals the combination is applied to and at what level the resulting proxy is defined, for example at token, span, entity, or full-response level.

For combination rule, works differ in whether the response-level estimate is refined sequentially as token-level evidence accumulates~\citep{gao2025flue,vazhentsev2025uncertainty}, or obtained by directly pooling token- or trace-level quantities once the relevant sequence-level evidence has been assembled~\citep{moslonka2025learned,huang2025reppl}, with both strategies producing response-level proxies used in downstream uncertainty estimation. For aggregation scope, the key variation is whether the response-level proxy pools over all positions, over a selected subset of salient positions, or over an intermediate unit such as an entity or span. Some methods explicitly restrict scope by selecting uncertainty-salient or reasoning-critical tokens before aggregation~\citep{zhou2025can,zhang_cot-uq_2025}, while others first define a local unit such as an entity and aggregate token-level scores within that boundary~\citep{yeh2025halluentity}.

A useful boundary case is a method whose main contribution lies in reading out a proxy from internal traces or features rather than in pooling explicitly defined local signals; such work belongs in readout transformation (P1.2.2). If the signals being consolidated are parallel within-request realisations rather than positional indices, the case belongs in multi-sample consolidation (P1.2.3).

\paragraph{Readout from internal traces (P1.2.2)}
\label{sec:p122}

P1.2.2 covers within-request transformation in which a new uncertainty proxy is constructed directly from model-internal features, such as hidden states, attention structure, or circuit-level signals, within a single model-facing request.  The defining feature is the construction of a proxy from model-internal states rather than from already-defined local uncertainty signals.
The salient transformation is the construction of \(s_{\mathrm{read}}\) from model-internal states available within the same request.

Two readout forms are especially prevalent. \emph{Mediated readout} constructs the proxy via a learned auxiliary component that maps internal states or traces to an uncertainty or correctness estimate, introducing a trained component between the internal states and the proxy. \emph{Direct readout} constructs the proxy by explicitly transforming intrinsic intermediate signals, such as hidden-state relations or intermediate predictive distributions, without a separate learned predictor.

For mediated readout, auxiliary predictors over internal states or traces have been shown to produce reliable scalar uncertainty or correctness estimates without requiring multiple samples~\citep{dakhmouche2025can,ghasemabadi2025can,zhou2025can}. Relatedly, probabilistic value heads can map hidden states to uncertainty-parameterized output distributions directly~\citep{lou2024uncertainty}. For direct readout, intrinsic intermediate signals such as softmax-response confidence and hidden-state saturation can be used as uncertainty proxies for early-exit decisions~\citep{schuster2022confident}. The key variation across both forms is whether a learned auxiliary component mediates the mapping from internal states to the proxy or whether the proxy is obtained directly from intermediate signals already available within the forward pass.

A useful boundary case is a method whose main contribution lies in pooling already-defined local signals rather than constructing a proxy from model-internal states; such work belongs in aggregation transformation (P1.2.1). If the main contribution instead lies in consolidating signals across parallel within-request realisations, it belongs in multi-sample consolidation (P1.2.3).

\paragraph{Consolidation from multiple samples (P1.2.3)}
\label{sec:p123}

P1.2.3 covers within-request transformation in which multiple stochastic realisations generated within a single model-facing request are consolidated into one uncertainty proxy. The defining move is consolidation across parallel within-request realisations, with cross-sample variation providing the evidence for the resulting proxy.
For sample-level signals \(\{s^{(j)}\}_{j=1}^{M}\) from \(M\) within-request realisations, consolidation yields a within-request uncertainty statement \(u_{\mathrm{ms}}=\langle s_{\mathrm{ms}},o,d,r,(a,c)\rangle\), where the salient transformation is \(\{s^{(j)}\}_{j=1}^{M} \to s_{\mathrm{ms}}\).

Two aspects of multi-realisation uncertainty propagation recur. \emph{Sampling strategy} refers to how multiple realisations are produced within the request boundary, whether through stochastic decoding across sampled generations or through randomized Monte Carlo-style perturbation schemes during inference. \emph{Variation retention} refers to how much of the resulting uncertainty structure survives into the final proxy: whether it is reduced to a scalar or response-level estimate, or retained first in a richer signal form such as a probability interval or credal-set style bound.

For sampling strategy, some methods obtain within-request realisations by stochastic decoding and compare the resulting sampled generations~\citep{huang2025reppl}, while others induce Monte Carlo-style realisations through weight perturbation or layer-wise internal randomization during inference~\citep{zhangtokur,gao2025flue}. For variation retention, most approaches reduce multi-realisation evidence to a scalar or response-level estimate, but recent work also represents higher-order uncertainty through probability intervals and credal-set style bounds before any later scalar summarization~\citep{yang2026verbalizing}. The key variation across both aspects is therefore not only how realisations are generated, but also whether uncertainty is scalarized immediately or first retained in interval- or set-valued form.

A useful boundary case arises when the consolidated signals are produced by controlled interventions: such work belongs here when the sampling strategy and variation retention are the main analytic object, and in intervention-conditioned transition (P1.1.3) when the intervention design and induced divergence are the primary contribution. When the signals being combined are positional indices within a single trace rather than parallel realisations, the case belongs in aggregation transformation (P1.2.1).

\subsubsection{Uncertainty-conditioned inference control (P1.3)}
\label{sec:p13}

\begin{wrapfigure}{r}{0.56\textwidth}
    \centering
    \includegraphics[width=\linewidth]{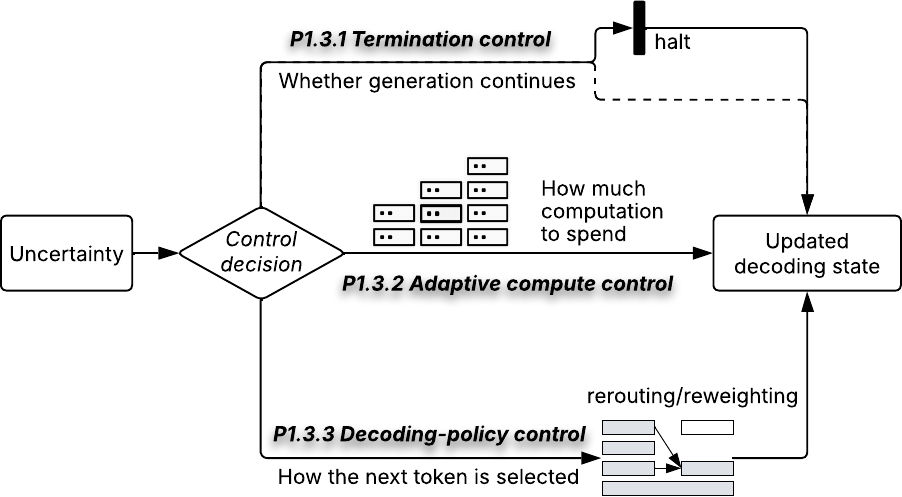}
    \caption{Uncertainty-conditioned inference control (P1.3)}
    \label{fig:p1.3}
\end{wrapfigure}

P1.3 covers within-request propagation in which an uncertainty signal is consumed to control how computation proceeds within a single model-facing request (See Figure \ref{fig:p1.3}). Where P1.1 traces how a signal evolves along an ordered sequence of internal positions and P1.2 asks how it is re-expressed into a proxy of a different type or scope, P1.3 concerns how a signal steers the subsequent trajectory of inference: the signal does not change form or accumulate evidence but determines what happens next. The statement remains at \(r_{\mathrm{int}}\) throughout.

The salient feature is the uncertainty signal is consumed with decision role \(d_{\mathrm{ctrl}}\), in contrast to P1.1 and P1.2 where the typical role is \(d_{\mathrm{diag}}\). Here the signal directly determines what happens next, for example whether generation continues, how much internal computation is allocated, or how the next token is selected. Propagation occurs through the resulting change in computation path rather than through a change in signal form.

What propagates in P1.3 is uncertainty as a within-request control input whose effect is realised through a changed computation path: gating whether generation continues (\hyperref[sec:p131]{P1.3.1}), adjusting how much internal computation each step receives (\hyperref[sec:p132]{P1.3.2}), or modifying the rule by which the next token is selected (\hyperref[sec:p133]{P1.3.3}).

\paragraph{Termination control (P1.3.1)}
\label{sec:p131}

P1.3.1 covers cases in which an uncertainty statement available during decoding is consumed to decide whether generation should continue or halt within the same model-facing request. The controlled quantity is the number of generation steps taken before the response is finalised: the signal determines whether remaining decoding steps are executed at all, rather than shaping how each step proceeds.

The central pattern in this subtype is \emph{uncertainty-gated stopping}: a within-request uncertainty proxy serves as an online halting criterion during decoding, foreclosing further computation rather than redirecting it. For example, in one documented instance, a local confidence/uncertainty proxy derived from the next-token distribution halts low-confidence reasoning traces early during parallel generation~\citep{fu2025deep}, with the halting decision itself constituting the propagation event.

A useful boundary case is control over how much internal computation is spent per token rather than whether generation continues at all; such work belongs in adaptive compute control (P1.3.2).

\paragraph{Adaptive compute control (P1.3.2)}
\label{sec:p132}

P1.3.2 covers within-request propagation in which an uncertainty signal is consumed to adjust how much internal computation is allocated before the current generation step is finalised. The controlled quantity is the local compute budget: an exit depth, an internal-step budget, or a choice of which active prefix to continue.

Two directions of this control pattern recur. \emph{Compute reduction} uses an internally derived uncertainty signal to select an earlier exit point, finalising the current token at a shallower depth and reducing computation while allowing decoding to proceed. \emph{Compute expansion} uses an internally derived uncertainty signal to allocate additional test-time computation, or to reallocate budget across active prefixes, when further computation is predicted to be beneficial.

For compute reduction, confidence-based early exit has been shown to reduce within-request computation substantially without degrading response quality~\citep{schuster2022confident}. For compute expansion, internally derived uncertainty signals about likely downstream success and remaining cost  have been used to adaptively continue, branch, prune, or pause active trajectories, thereby concentrating computation where the current partial generation appears less settled or more worth further exploration~\citep{manvi2025zero}. Across both directions, the propagated signal steers resource allocation within the request.

A useful boundary case is control over whether generation continues at all; such work belongs in termination control (P1.3.1). Control over next-token selection rather than compute allocation belongs in decoding-policy control (P1.3.3).

\paragraph{Decoding-policy control (P1.3.3)}
\label{sec:p133}

P1.3.3 covers within-request propagation in which an uncertainty signal is consumed to modify the rule by which the next token is selected within a single model-facing request. Typical actions include adjusting sampling parameters, reweighting candidate tokens under an auxiliary objective, or switching between decoding modes.

Two scopes of policy adjustment recur in this subtype. \emph{Step-level policy adjustment} acts at the current decoding step by reweighting or recombining candidate tokens in response to the uncertainty signal. \emph{Search-level policy adjustment} responds to persistent uncertainty by triggering a broader change in decoding mode or search procedure, while remaining within the same request.

Both scopes are documented in the literature. For step-level policy adjustment, energy-based uncertainty signals derived from model logits have been used to dynamically weight model contributions when forming the token-selection distribution during contrastive decoding~\citep{lee2025uncertainty}. For search-level policy adjustment, token-level uncertainty estimates have been used to trigger online revision of the decoding rule, first through probability recalibration toward context-consistent candidates and, when uncertainty persists, through escalation to broader search within the same request~\citep{yarie2024mitigating}. Across both scopes, the uncertainty signal reshapes how the next token is chosen rather than how many tokens are generated or how much computation each token receives.

A useful boundary case is control over compute allocation per token rather than token selection; such work belongs in adaptive compute control (P1.3.2).

\subsection{System-level uncertainty propagation across the technical system (P2)}
\label{sec:cross-component}

\begin{figure}
    \centering
    \includegraphics[width=0.75\linewidth]{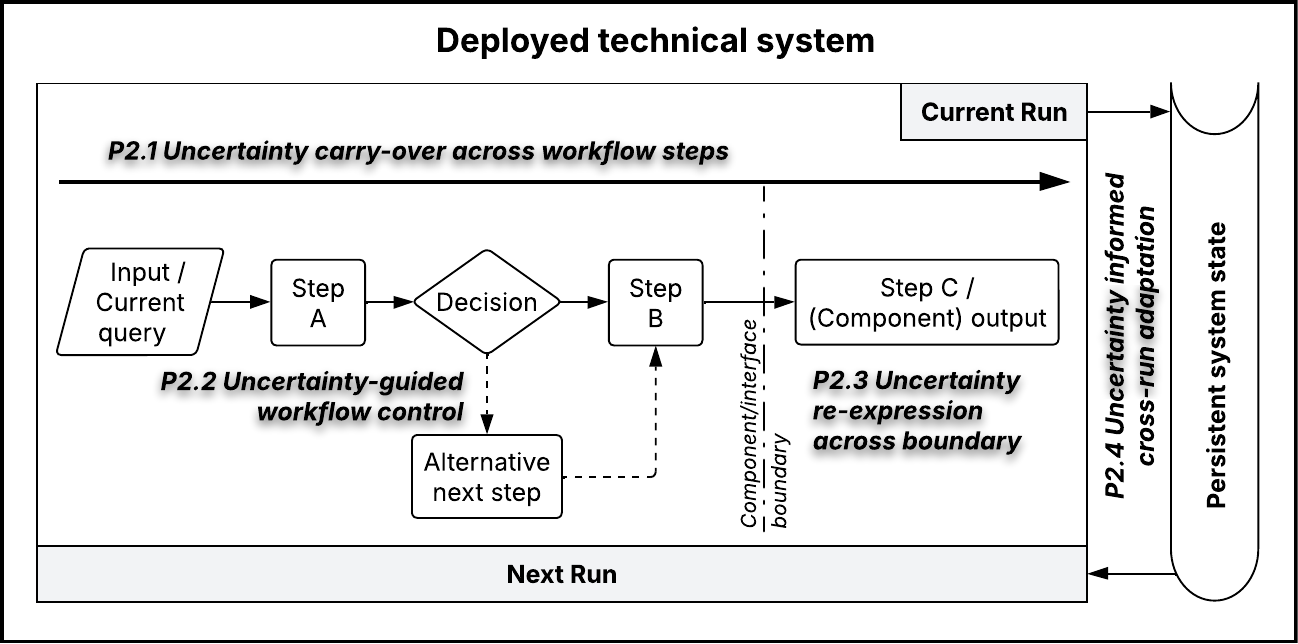}
    \caption{Schematic overview of system-level uncertainty propagation in P2.}
    \label{fig:p2_overview}
\end{figure}

P2 covers uncertainty propagation that begins when an uncertainty signal leaves the model-internal locus \(r_{\mathrm{int}}\) and becomes available in system-held state \(r_{\mathrm{sys}}\) for consumption by downstream technical components (see Figure \ref{fig:p2_overview}). At this point uncertainty is no longer an internal property of model computation; it becomes part of the system's operational state and control logic. This distinguishes P2 from P1, which remains within a single model-facing request, and from P3, where uncertainty reaches human-facing or assurance-artefact loci beyond the deployed technical system. At each such boundary, any field of the uncertainty statement \(u = \langle s, o, d, r, (a,c) \rangle\) may be preserved, transformed, or silently re-specified, and it is this potential for field-level change at component boundaries that makes P2 propagation analytically distinct from within-request propagation.

P2 groups recurring propagation patterns by how uncertainty is carried, reused, and re-expressed within the deployed technical system: carried forward across workflow steps (\hyperref[sec:p21]{P2.1}), consumed to determine the next system action (\hyperref[sec:p22]{P2.2}), re-expressed across a component boundary (\hyperref[sec:p23]{P2.3}), or retained beyond a run to adapt future system behaviour (\hyperref[sec:p24]{P2.4}).

\subsubsection{Uncertainty carried across workflow steps (P2.1)}
\label{sec:p21}

\begin{wrapfigure}{r}{0.45\textwidth}
    \centering
    \includegraphics[width=\linewidth]{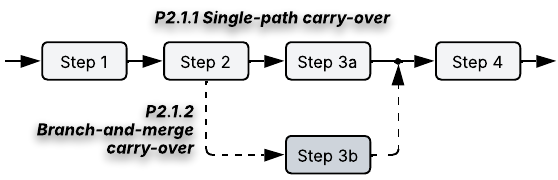}
    \caption{Uncertainty carried across workflow steps (P2.1)}
    \label{fig:p2.1}
\end{wrapfigure}

P2.1 covers system-level propagation in which an uncertainty signal is carried forward across successive workflow steps within the same system run (see Figure \ref{fig:p2.1}). The defining feature is continuity of the realised execution trace: the run proceeds along its current path, and propagation is analysed through how the signal is preserved and updated as it moves from one step to the next as part of workflow state.

For two successive workflow steps within the same run, let \(u\) and \(u'\) denote the earlier and later contextualised uncertainty statements, written descriptively as \(u \to u'\). In P2.1, the central question is not which path is selected, but how uncertainty already on the realised path is carried forward and updated across step-to-step execution. In many cases the dominant visible update is in the uncertainty signal \(s \to s'\), although other fields may remain stable or shift locally as the statement is reused by the next step within the same run.

What propagates in P2.1 is an uncertainty signal carried as part of within-run workflow state, accumulating or updating across steps rather than being estimated independently at each stage. This carry-forward may proceed along a single continuing path through a non-branching segment (\hyperref[sec:p211]{P2.1.1}), or across multiple branch-local paths whose uncertainty is later brought together at a merge point (\hyperref[sec:p212]{P2.1.2}).

\paragraph{Single-path carry-over in non-branching segments (P2.1.1)}
\label{sec:p211}

P2.1.1 covers non-branching segments of a system run in which an uncertainty signal is carried forward across successive workflow steps. The defining feature is later uncertainty is analysed as depending on earlier carried-over uncertainty, rather than as a fresh estimate attached independently to each stage.
For successive workflow steps \(i\) and \(i+1\), we write \(u_i \to u_{i+1}\), where the salient update is often \(s_i \to s_{i+1}\).

The recurring pattern is \emph{sequential history dependence}, in which run-level uncertainty depends on accumulated stepwise history rather than independent per-step estimates. This appears most clearly in methods that decompose per-step uncertainty into local and inherited components~\citep{duan2025uprop}, and more broadly in trajectory-level approaches that derive run-level confidence from cross-step uncertainty traces and process dynamics~\citep{zhao2025uncertainty,zhang2026agentic}. What these works share is that the sequential dependency structure is itself the primary object of analysis rather than a byproduct of the estimation procedure.

A useful boundary case is a method that reports step-indexed uncertainty values but treats each step independently rather than as part of a carried-forward trace; such work does not fit P2.1.1 regardless of how steps are indexed. If uncertainty is instead propagated along diverging branch-local paths before merge, the case belongs in P2.1.2. If the main contribution lies in re-expressing for downstream technical reuse, it belongs in P2.3.

\paragraph{Multi-path carry-over in branch-and-merge segments (P2.1.2)}
\label{sec:p212}

P2.1.2 covers branch-and-merge segments of a system run in which uncertainty signals are carried forward on multiple explicit workflow branches before later merge. The defining feature is branch-local carry-forward with merge: uncertainty propagated on separate branches is later compared or combined when the workflow returns to a common continuation, producing a single post-merge signal that the continuation acts on.

Branch reconciliation in this subtype takes two forms that recur across the surveyed literature. \emph{Merge by aggregation} combines branch-level confidence or uncertainty-sensitive scores into a post-merge decision state, as in confidence-weighted aggregation over multiple sampled reasoning paths~\citep{razghandi2025cer}. \emph{Merge by selection} instead commits to one surviving candidate by ranking branch- or sample-local reliability signals, for example by selecting the least suspicious response from a sampled set~\citep{phillips2025geometric}. The key distinction is whether uncertainty-sensitive information contributes to a combined post-merge score or is used only to retain one surviving branch.

A useful boundary case is when alternatives exist only within a single model-facing request rather than as explicit workflow branches; such work belongs in multi-sample consolidation (P1.2.3). When uncertainty determines whether branches are created, expanded, or pruned rather than how they are reconciled at merge, the case belongs in P2.2.

\subsubsection{Uncertainty-guided workflow control (P2.2)}
\label{sec:p22}

P2.2 covers system-level propagation in which an uncertainty signal is consumed to determine a subsequent system action. The defining feature is consumption of the signal in a \(d_{\mathrm{ctrl}}\) role at a workflow decision point: rather than being carried forward along the current path, the signal influences what happens next. This distinguishes P2.2 from P2.1, where uncertainty is carried along an already chosen execution path, and from P2.3, where the signal is reformulated for a receiving component rather than consumed to determine an action.

At a workflow decision point, the salient update is \(d \to d_{\mathrm{ctrl}}\): an uncertainty signal previously in a diagnostic role is consumed to determine the next system action. The reuse locus or attribution-consumer pair may also shift at the decision point, depending on whether the control action crosses a component or actor boundary.

What propagates in P2.2 is an uncertainty signal that crosses from a diagnostic role into a control role, triggering a change in system execution path rather than continuing along the current one. P2.2 groups three recurring forms of this control consumption: planning and search control (\hyperref[sec:p221]{P2.2.1}), routing across strategy or model tiers (\hyperref[sec:p222]{P2.2.2}), and assistance or escalation gating (\hyperref[sec:p223]{P2.2.3}).

\begin{wrapfigure}{r}{0.55\textwidth}
    \centering
    \includegraphics[width=\linewidth]{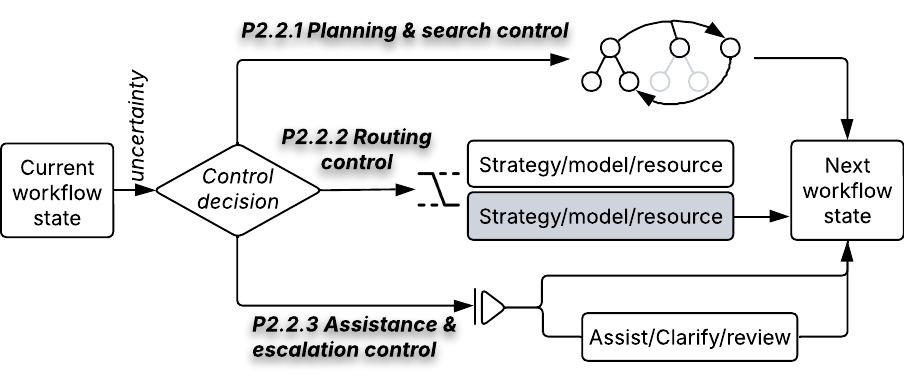}
    \caption{Uncertainty-guided workflow control (P2.2)}
    \label{fig:p2.2}
\end{wrapfigure}

\paragraph{Planning and search control (P2.2.1))}
\label{sec:p221}

P2.2.1 covers system-level propagation in which an uncertainty signal is consumed within a planning or search loop to control how exploration proceeds. The defining feature is in-loop exploration control: rather than being applied only after a completed candidate has been produced, uncertainty is reused to decide which search state, partial plan, or intermediate candidate to pursue next, whether to prune or backtrack, and whether to gather more evidence before proceeding.

Two patterns instantiate this subtype. \emph{Exploration steering} directs the loop toward more promising regions by using uncertainty to influence which partial plans, search states, or intermediate reasoning states are expanded or pruned next. \emph{Information-augmenting control} suspends forward progress by using uncertainty to trigger an additional information-bearing step before the loop continues, such as querying for further evidence or introducing corrective guidance. The distinction between them is directional: exploration steering redirects the trajectory, whereas information-augmenting control pauses it and augments the state available for continuation.
The object of concern also differs: exploration steering typically concerns an output artefact (\(o_{\mathrm{art}}\)), namely whether a partial plan or candidate state is adequate to pursue, while evidence-seeking control typically concerns a decision prerequisite (\(o_{\mathrm{input}}\)), namely whether sufficient information is available to proceed.

Both patterns appear in the surveyed literature, though more sparsely than the subtypes that follow. Exploration steering appears in systems that use uncertainty-aware value estimates or uncertainty-sensitive search scores to guide plan expansion, candidate retention, or pruning during search~\citep{dengplanu,yu2025uncertainty}. Information-augmenting control appears in systems that use uncertainty or expected information gain to trigger additional querying or information gathering before proceeding~\citep{choudhury2025bed,hu2024uncertainty}, as well as in systems that use rising uncertainty to interrupt an ongoing reasoning trajectory and introduce corrective reasoning clues before generation resumes~\citep{yin2024reasoning}. In both forms, uncertainty is reused inside the planning or search loop itself to control how exploration proceeds.

A useful boundary case is a branch-and-merge structure where the main concern is reconciliation of branch-local uncertainty at merge rather than control within the loop; such work belongs in P2.1.2. When uncertainty routes among pre-existing model or strategy tiers rather than controlling exploration within a loop, the case belongs in P2.2.2.

\paragraph{Routing control across internal system tiers (P2.2.2)}
\label{sec:p222}

P2.2.2 covers system-level propagation in which an uncertainty signal is consumed to select among pre-existing technical routes within the deployed system. The defining feature is route selection among already available internal alternatives: the signal determines which route to take rather than how to explore within an evolving search process.

The recurring pattern is \emph{uncertainty-conditioned route selection}. Three routing targets are distinguished in this subtype. \emph{Tiered model or verification routing} forwards a query to a stronger model or defers to a stronger verification tier when the low-cost route is not sufficiently trustworthy. \emph{Resource-tier routing} redirects inference from a local or on-device model to a remote model when local uncertainty exceeds a threshold. \emph{Reasoning-strategy routing} switches from a direct generation path to a more elaborate internal reasoning procedure under elevated uncertainty.

Tiered model or verification routing appears in systems that tune confidence thresholds for LLM cascades or selectively defer from weak to strong verification when low-cost signals are insufficiently reliable~\citep{zellinger2025rational,kiyani2026trust}. Resource-tier routing appears in hybrid on-device and remote inference systems that escalate from a local small model to a remote large model when local uncertainty exceeds a threshold~\citep{oh2025uncertainty}. Reasoning-mode routing appears in systems that invoke a more elaborate reasoning path only when uncertainty exceeds a threshold, while otherwise remaining on a direct generation path~\citep{zhu2025uncertainty}. Across all three targets, uncertainty gates access to a stronger or more resource-intensive route rather than steering exploration within a search loop.

A useful boundary case is uncertainty that controls exploration within an evolving search process rather than selecting among pre-existing routes; such work belongs in planning and search control (P2.2.1). When the main propagation event lies in re-expressing the signal at the boundary created by the chosen route, it belongs in P2.3.

\paragraph{Assistance and escalation control (P2.2.3)}
\label{sec:p223}

P2.2.3 covers system-level propagation in which an uncertainty signal determines whether the system may continue autonomously or must first enter an assistance or escalation channel. The defining feature is qualified autonomy: under current uncertainty, autonomous continuation is withheld pending clarification, confirmation, or additional input. The control question is not which autonomous route to take next but whether autonomous progress is justified at all.
The object of concern is typically a decision prerequisite (\(o_{\mathrm{input}}\)): whether the available action space, parameter specification, or contextual information is sufficient to support confident autonomous continuation.

The recurring pattern is an \emph{assistance gate}: uncertainty is used at a workflow decision point to decide whether additional input is required before the next step. Two main trigger forms exist. \emph{Next-action ambiguity} arises when multiple plausible next actions remain and the system cannot confidently commit to one autonomously. \emph{Argument ambiguity} arises when inputs required for the next action remain ambiguous, incomplete, or insufficiently resolved for safe execution.

Next-action ambiguity appears in systems that ask for help when calibrated uncertainty leaves multiple plausible next actions~\citep{renrobots}. Argument ambiguity appears in systems that trigger clarification when tool arguments required for the next action remain unresolved~\citep{suri2025structured}. In both cases, the propagation event is the workflow decision to open the assistance or escalation channel, not any later uptake by an external actor.

A useful boundary case is uncertainty that controls exploration within a planning or search loop rather than gating autonomous continuation; such work belongs in P2.2.1. When uncertainty selects among pre-existing autonomous routes rather than suspending autonomy, it belongs in P2.2.2. Any uptake of the escalated signal by a human, organisation, or governance process belongs in P3.

\subsubsection{Uncertainty re-expression across component boundaries (P2.3)}
\label{sec:p23}

\begin{wrapfigure}{r}{0.4\textwidth}
    \centering
    \includegraphics[width=0.95\linewidth]{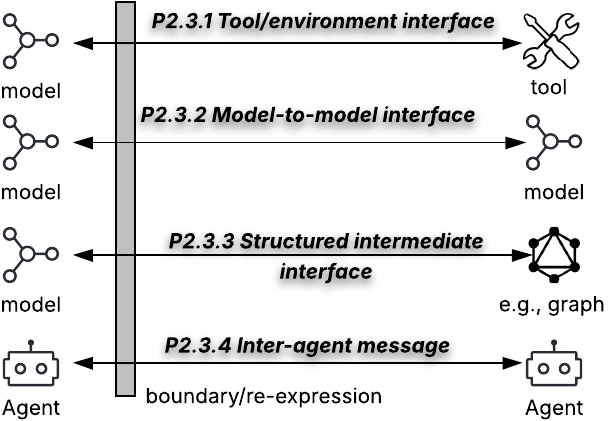}
    \caption{Uncertainty re-expression (P2.3)}
    \label{fig:p2.3}
\end{wrapfigure}

P2.3 covers system-level propagation in which an uncertainty signal is re-expressed for consumption by a downstream technical component across a system boundary (see Figure \ref{fig:p2.3}). The defining feature is re-expression at a component boundary: the signal is encoded, attached to an artefact, or passed in a boundary-compatible form as the consumer and attribution shift from one component to another. This distinguishes P2.3 from P2.1, where uncertainty is carried along an already chosen execution path, and from P2.2, where uncertainty determines the next system action rather than being reformulated for a receiving component.

At a component boundary, the most common updates are \(s \to s'\) and \((a,c) \to (a',c')\): the signal changes form to suit the receiving component, and the consumer shifts as the signal crosses the boundary. In some cases \(s\) remains stable while only \((a,c)\) shifts. The reuse locus remains at \(r_{\mathrm{sys}}\) throughout. The object of concern \(o\) and decision role \(d\) may be preserved or silently re-specified at the boundary; silent re-specification of \(o\) or \(d\) is the primary failure mode this subsection addresses. The decision role ranges from \(d_{\mathrm{ctrl}}\) when the re-expressed signal steers a downstream action to \(d_{\mathrm{diag}}\) when it informs downstream assessment.

What propagates in P2.3 is an uncertainty signal in a boundary-adapted form whose object of concern and decision role may shift as it crosses from one component to another. P2.3 groups four recurring forms of boundary re-expression: tool and environment interface re-expression (\hyperref[sec:p231]{P2.3.1}), model-to-model interface re-expression (\hyperref[sec:p232]{P2.3.2}), structured intermediate re-expression (\hyperref[sec:p233]{P2.3.3}), and inter-agent message re-expression (\hyperref[sec:p234]{P2.3.4}).

\paragraph{Tool and environment interface re-expression (P2.3.1)}
\label{sec:p231}

P2.3.1 covers system-level propagation in which uncertainty at a tool or execution-environment boundary is re-expressed in an interface-specific form that downstream technical components can use. The boundary may be a simple API call or a higher-level callable capability, such as an agent skill~\citep{jiang2026sok}. The defining feature is interface-level encoding: uncertainty is encoded in terms of the specific variables exposed at the tool boundary rather than treated as a generic model-side confidence signal.

The recurring pattern here is \emph{interface uncertainty encoding}. Two forms correspond to the two sides of a tool interaction. \emph{Pre-invocation argument encoding} concerns the input-side object of concern \(o_{\mathrm{input}}\): whether the tool call is sufficiently specified to execute. In this form, underspecified or ambiguous tool arguments are expressed as structured uncertainty over tool-call candidates and parameter domains, so that the system can repair or clarify the call before invocation. \emph{Post-invocation result composition} concerns the output-side object of concern \(o_{\mathrm{art}}\): how uncertainty in a returned observation or tool result contributes to the reliability of the subsequent system artefact. In this form, uncertainty from the tool output is combined with uncertainty in the LLM-generated answer to estimate the uncertainty of the overall tool-augmented response.
The decision role is typically \(d_{\mathrm{ctrl}}\): the encoded signal steers whether the call proceeds, is repaired, or is clarified on the input side, and whether the result is accepted or flagged on the output side.

For pre-invocation argument encoding, it appears in systems that maintain structured beliefs over tool-call candidates and parameter domains, using those beliefs to decide whether to clarify or execute~\citep{suri2025structured}. Post-invocation result composition appears in systems that model the tool-calling process through tool-call, tool-output, and final-answer variables, then combine tool predictive uncertainty with LLM answer uncertainty to estimate the reliability of the combined response~\citep{lymperopoulos2025tools}. Across both forms, the propagation move is not merely that tools introduce uncertainty, but that boundary-specific uncertainty is recoded into a system-level signal that can constrain later execution, interpretation, or reliability assessment.

A useful boundary case is uncertainty used to decide whether to invoke, retry, or clarify before proceeding; such work belongs in P2.2 since the signal is consumed to determine the next action rather than encoded for downstream use. Any uptake of the re-expressed signal by a human, organisation, or physical-world process belongs in P3.

\paragraph{Model-to-model re-expression (P2.3.2)}
\label{sec:p232}

P2.3.2 covers system-level propagation in which an uncertainty signal produced at one model stage is consumed across a model boundary by another model stage, such as a verifier, judge, critic, or value model. The defining feature is meaning-sensitive consumption: the receiving stage consumes the signal as part of its own assessment, and the signal's object of concern, decision role, and consumer must remain explicit so that a local signal is not treated as broader evidence than it actually provides.

The recurring pattern is \emph{meaning-preserving model handoff}. The propagation concern is whether the signal retains its intended scope when it crosses from the producing to the receiving model stage. A local evaluator signal reused without explicit scoping can be treated downstream as if it were ground truth or as if it supported a broader reliability claim than the producing stage actually justifies. This pattern appears in systems that use uncertainty-aware value models to assess partial reasoning paths whose assessments are subsequently reused in search or generation~\citep{yu2025uncertainty}, and in evaluation frameworks where judge outputs are reported with explicit uncertainty and scope conditions rather than forwarded as raw scores~\citep{lee2025correctly}. Across these cases, the central propagation issue is not the numerical transfer of a score alone, but the preservation of its assessment scope when another model stage acts on it.

A useful boundary case is uncertainty that decides whether another model stage is invoked, but is not itself consumed by that stage as an assessment signal. Confidence-based cascade deferral and local-uncertainty-based remote verification (e.g., \citep{zellinger2025rational,oh2025uncertainty}) are routing cases: uncertainty selects a model tier or execution location, so they belong in routing control (P2.2.2). P2.3.2 is narrower: the uncertainty-bearing signal must cross the model boundary and be interpreted by the receiving model-facing component. If the signal is instead embedded in a structured intermediate such as a graph or probabilistic program, the case belongs in structured intermediate re-expression (P2.3.3).

\paragraph{Structured intermediate re-expression (P2.3.3)}
\label{sec:p233}

P2.3.3 covers system-level propagation in which uncertainty is re-expressed through a structured intermediate, such as a graph or probabilistic program, as it crosses a component boundary. The defining feature is: the receiving component must interpret the uncertainty together with the structured variables, nodes, edges, triples, clauses, or dependencies through which it is represented.
The object of concern is tied to a specific structured element such as a node, edge, clause, or program component rather than to the intermediate as a whole (\(o_{\mathrm{art}}\)), and the decision role is typically \(d_{\mathrm{ctrl}}\): the receiving component uses the structure-bound signal to reason over or execute the intermediate, with the signal's element attachment determining which downstream actions it can support.

The recurring pattern is \emph{structure-bound uncertainty encoding}: uncertainty is not forwarded as a standalone score, but is bound to a structured intermediate that determines what the signal qualifies and how it can be propagated or inferred over. Two forms instantiate this pattern. \emph{Graph-mediated encoding} represents uncertainty through graph structure, for example by deriving uncertainty from graph embeddings or by propagating and calibrating uncertainty over graph elements and relations. \emph{Probabilistic-program encoding} re-expresses uncertainty in an executable probabilistic representation, preserving both the uncertain quantities and the dependency structure for downstream inference.

Graph-mediated encoding appears in systems that construct knowledge or semantic graphs from generated text and use the resulting graph structure to estimate, propagate, or calibrate uncertainty for hallucination detection and long-text reliability assessment~\citep{yuan2025kg,chen2025enhancing}. Probabilistic-program encoding appears in systems that translate probabilistic information expressed in text into executable symbolic or probabilistic programs, allowing a downstream solver to carry the encoded uncertainty through inference~\citep{nafar2025reasoning}. Across both forms, the propagation risk is structural flattening: the uncertainty value may survive, but the structured attachment that determines what it qualifies may be lost.

A useful boundary case is uncertainty passed directly from one model-facing stage to another without structure-bound attachment; such work belongs in model-to-model re-expression (P2.3.2). When uncertainty controls the next route or action rather than being encoded in a structured intermediate, the case belongs in P2.2.

\paragraph{Inter-agent re-expression (P2.3.4)}
\label{sec:p234}

P2.3.4 covers system-level propagation in which agents in a multi-agent protocol exchange uncertainty-bearing messages. The defining feature is protocol-mediated re-expression: an agent encodes local uncertainty in a message form that another protocol participant can inspect and act on, rather than leaving uncertainty implicit in the generated content alone.
The object of concern is typically an output artefact (\(o_{\mathrm{art}}\)), namely the reliability or confidence of an agent's contribution, and the decision role is \(d_{\mathrm{comm}}\): uncertainty is disclosed to shape how other agents weight, filter, or update their local state in response.

The recurring pattern is \emph{uncertainty-disclosing message exchange}: local uncertainty is re-expressed in a protocol-legible form, so that receiving agents can use it when interpreting, weighting, or revising contributions. Two forms instantiate this pattern. \emph{Confidence-weighted exchange} attaches a confidence or uncertainty estimate to an agent contribution, allowing receivers to weight peer inputs during later aggregation or response generation. \emph{Belief-state disclosure} communicates richer uncertainty-bearing state, such as confidence, unresolved assumptions, remaining unknowns, or explicit belief proposals, allowing receivers to update their own belief state or final assessment.

Both forms are documented in the surveyed literature. Confidence-weighted exchange appears in debate systems that attach uncertainty-derived confidence values to agent responses so that later agents can adjust how much influence peer contributions receive~\citep{yoffe2024debunc}. Belief-state disclosure appears in debate protocols where agents explicitly report confidence and remaining uncertainties before final judgment~\citep{liu2025uncertainty}, and in decentralized belief-propagation frameworks where agents exchange belief proposals that listeners evaluate and incorporate through local update rules~\citep{hayashi2025decentralized}. Across these cases, uncertainty becomes part of the interaction protocol itself: it is disclosed in message form and shapes how other participants interpret, weight, or revise the information they receive.

A useful boundary case is a single signal passed between model stages without protocol-mediated exchange; such work belongs in model-to-model re-expression (P2.3.2). Ordinary multi-agent debate belongs here only when uncertainty is explicitly encoded in the message or update protocol.

\subsubsection{Cross-run uncertainty propagation through system adaptation (P2.4)}
\label{sec:p24}
\begin{wrapfigure}{r}{0.45\textwidth}
    \centering
    \includegraphics[width=\linewidth]{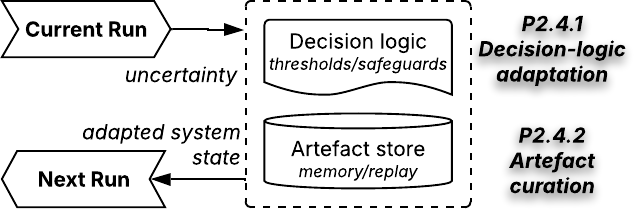}
    \caption{Cross-run adaptation (P2.4)}
    \label{fig:p2.4}
    \vspace{-10pt}
\end{wrapfigure}

P2.4 covers system-level propagation in which an uncertainty signal observed in one run is retained in persistent system state and later changes how subsequent runs proceed, including both deployed runtime pipelines and iterative training or alignment pipelines where uncertainty shapes what future executions inherit. The defining feature is cross-run adaptation: uncertainty is not only used within the current execution, but recorded in a form that survives run completion and influences later executions. This distinguishes P2.4 from P2.1--P2.3, which concern propagation within a single run, and from P3, where persistent uncertainty records are taken up by human, organisational, or governance processes outside the deployed technical system.

At the end of a run, an uncertainty signal with decision role \(d_{\mathrm{adapt}}\) contributes to an update in persistent system state that later runs inherit. The signal may be stored directly, aggregated across runs, or translated into an update rule; what matters is that the conditions under which later runs execute are changed as a result. The reuse locus remains at \(r_{\mathrm{sys}}\) throughout.
Across both subtypes, the object of concern is operating conditions and controls (\(o_{\mathrm{op}}\)): whether the current evidence, safeguards, or operating conditions are adequate for reliable future executions. The subtypes differ in what is updated: persistent decision logic (\hyperref[sec:p241]{P2.4.1}) or the retained artefact pool (\hyperref[sec:p242]{P2.4.2}).

What propagates in P2.4 is an uncertainty signal operationalised through persistent system state: retained beyond a run and changing the technical conditions under which later runs proceed. P2.4 distinguishes two recurring forms of this cross-run effect: updates to persistent decision logic that governs future runs (\hyperref[sec:p241]{P2.4.1}), and curation of the artefact store from which future runs draw (\hyperref[sec:p242]{P2.4.2}).

\paragraph{Decision-logic adaptation (P2.4.1)}
\label{sec:p241}

P2.4.1 covers cross-run propagation in which uncertainty retained from earlier executions, rollouts, or validation runs is used to update persistent decision logic that governs later runs. The defining feature is policy adaptation across runs: the signal changes the rules by which future executions are trained, selected, validated, or escalated, rather than affecting only the current execution.

The recurring pattern is \emph{uncertainty-conditioned decision-logic revision}. Two forms instantiate this pattern. \emph{Learning-based policy adaptation} uses uncertainty in reward, preference, or trajectory quality to constrain learning updates, reweight training data, or penalize unreliable optimization targets, so that persistent policies adapt more cautiously when feedback is weak, ambiguous, or unreliable. \emph{Validation-based safeguard calibration} uses accumulated validation or monitoring evidence, such as output drift, consistency failure, or invariant violations, to configure future-run controls such as model-tier selection, acceptance gates, audit requirements, or escalation triggers.

Learning-based policy adaptation appears in alignment and agent-training methods that use reward-model or trajectory-level uncertainty to filter unreliable feedback, scale optimization strength, or shape training rewards~\citep{lou2024uncertainty,ciefadaptive,stoissertowards}. Validation-based safeguard calibration appears in deployment-oriented validation frameworks that use drift, consistency, or invariant-checking evidence to define persistent controls for later executions~\citep{khatchadourian2025llm}. Across both forms, uncertainty propagates by changing future decision logic: it is retained as evidence for how subsequent runs should be trained, accepted, reviewed, or constrained.

A useful boundary case is uncertainty consumed to determine the next action within the current run rather than to update persistent decision logic; such work belongs in uncertainty-guided workflow control (P2.2). If the main propagated object instead consists of retained data, memory entries, or trajectories rather than decision logic, the case belongs in artefact curation (P2.4.2).

\paragraph{Artefact curation (P2.4.2)}
\label{sec:p242}

P2.4.2 covers cross-run propagation in which uncertainty from earlier runs determines which artefacts remain available to later runs. The defining feature is \emph{artefact-pool carry-over}: uncertainty changes the persistent set of examples, labels, reward records, generated outputs, or synthetic supervision that future runs inherit, rather than revising the decision logic applied to those artefacts.

The recurring pattern is \emph{uncertainty-conditioned artefact curation}. Two forms instantiate this pattern. \emph{Coverage-aware selection} uses uncertainty together with representativeness or diversity constraints, so that the retained pool covers informative regions without concentrating only on isolated high-uncertainty outliers. \emph{Reliability-gated filtering} uses uncertainty as a direct quality gate, excluding artefacts whose labels, rewards, generated content, or synthetic supervision are too unreliable for future reuse.

Coverage-aware selection appears in active curation pipelines that use propagated uncertainty together with representativeness or diversity constraints to select samples for annotation, feedback, or later training~\citep{yu2023cold,liang2024actively}. For reliability-gated filtering, it appears in alignment and agent-training pipelines that use uncertainty to remove unreliable reward feedback, training pairs, generated summaries, or synthetic outputs from the artefact pool used by later runs~\citep{lou2024uncertainty,stoissertowards}. Across both forms, uncertainty propagates beyond local estimation by shaping what future executions inherit: it changes the available artefacts, not the decision logic that governs their use.

A useful boundary case is uncertainty consumed to determine an action within the current run rather than to curate a persistent artefact pool; such work belongs in uncertainty-guided workflow control (P2.2). If the main propagated effect instead lies in revising persistent decision logic rather than the artefact pool, the case belongs in decision-logic adaptation (P2.4.1).

\subsection{Socio-technical uncertainty propagation beyond the technical system (P3)}
\label{sec:sociotech_propagation}

\begin{figure}
    \centering
    \includegraphics[width=0.8\linewidth]{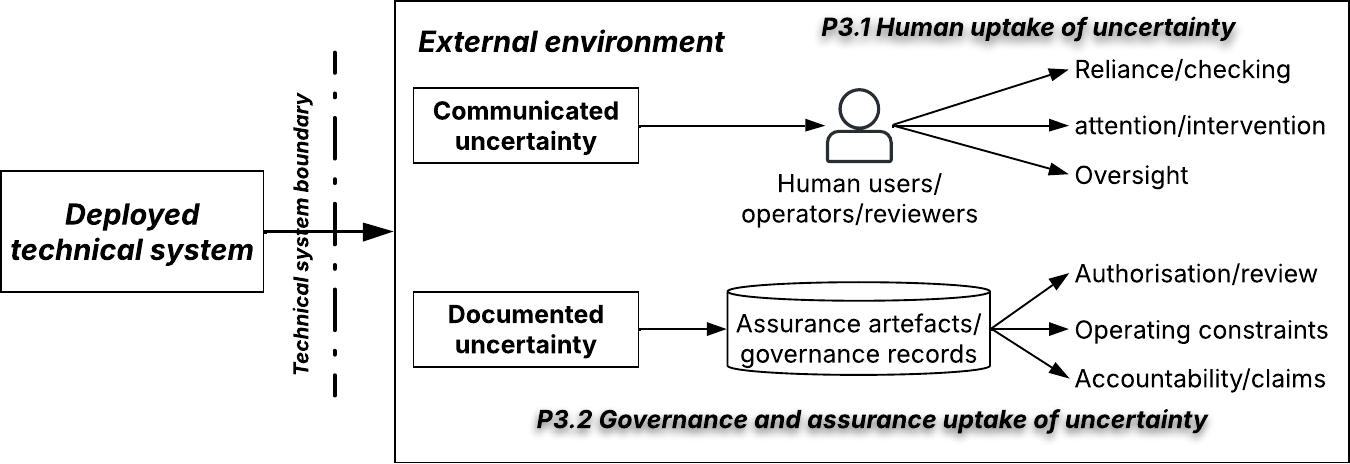}
    \caption{Schematic overview of socio-technical uncertainty propagation in P3.}
    \label{fig:p3_overview}
\end{figure}

P3 covers uncertainty propagation whose main effect is realised outside the deployed technical system (see Figure \ref{fig:p3_overview}). The defining feature is external uptake: uncertainty is presented beyond the system boundary and taken up by users, operators, organisations, or institutions, where it can shape reliance and verification behaviour, organisational decisions, and governance actions such as authorisation or operating constraints. This distinguishes P3 from P2, where uncertainty is reused inside the technical system for execution, coordination, or adaptation; note that if externally taken-up uncertainty is later fed back into the technical system, the outward segment belongs in P3 while the return path is best analysed as a new P2 propagation episode.

At the system boundary, an uncertainty signal leaves the model-internal or system-held locus and is re-expressed in a form suited for external consumption, written descriptively as \(u \to u'\) with \(r \to r'\) and \(r' \in \{r_{\mathrm{hum}},\, r_{\mathrm{assure}}\}\). In many cases the attribution-consumer pair also shifts as the signal is reformulated for external uptake. The reuse locus transition from \(r_{\mathrm{sys}}\) to \(r_{\mathrm{hum}}\) or \(r_{\mathrm{assure}}\) is the defining boundary crossing of P3.

P3 distinguishes two recurring forms of external uptake: human uptake of communicated uncertainty (\hyperref[sec:p31]{P3.1}), where the focus is on how people respond to disclosed uncertainty, and governance and assurance uptake of documented uncertainty (\hyperref[sec:p32]{P3.2}), where uncertainty is embedded in artefacts that support organisational and governance decisions.

\subsubsection{Human uptake of communicated uncertainty (P3.1)}
\label{sec:p31}

P3.1 covers socio-technical propagation in which an uncertainty signal is presented to humans and the phenomenon of interest is the resulting change in human reliance, verification, attention, or oversight behaviour. The defining feature is human uptake: uncertainty leaves the deployed technical system and becomes part of how people judge, verify, or act on system outputs.

What propagates in P3.1 is an uncertainty signal in a human-interpretable form, with decision role \(d_{\mathrm{comm}}\), whose effect is realised through impacted human behaviour rather than through system execution. The object of concern is typically an output artefact (\(o_{\mathrm{art}}\)): whether a specific system output is reliable or adequate enough to act on, though it may shift to a decision prerequisite (\(o_{\mathrm{input}}\)) when communicated uncertainty concerns whether the available information is sufficient to proceed.

The recurring pattern is \emph{uncertainty disclosure}: uncertainty produced inside the system is communicated in a human-interpretable form so that users can adjust reliance or direct verification and oversight. The form of disclosure varies across instantiations. In generated responses, natural-language expressions of uncertainty can reduce overreliance and improve task accuracy by prompting users to verify uncertain outputs~\citep{kim2024m}. In interactive analysis interfaces, uncertainty can be embedded in visual interface components that help analysts assess whether observed structures should be trusted before drawing conclusions~\citep{sevastjanova2025layerflow}. Across these cases, uncertainty propagates by becoming part of the human interpretation context: it shapes how people rely on, verify, or scrutinise system outputs rather than steering system execution directly.

A useful boundary case is a user-facing uncertainty display whose primary effect is triggering an automated system action such as retry or escalation; such work belongs in P2 since the main propagation effect is realised inside the technical system rather than through human response.

\subsubsection{Governance and assurance uptake of documented uncertainty (P3.2)}
\label{sec:p32_open}

P3.2 covers socio-technical propagation in which documented uncertainty is used as evidence in organisational governance. The defining feature is assurance and governance uptake: uncertainty no longer affects only immediate user reliance, but enters assurance artefacts, audit records, evaluation reports, or validation evidence that organisations use to justify, qualify, or constrain system use.
The object of concern is typically operating conditions and controls (\(o_{\mathrm{op}}\)): the question is not whether a specific output artefact is reliable, but whether the available evidence, safeguards, and operating conditions are sufficient to support the governance or authorisation decision at hand.
The key distinction from P3.1 is institutional uptake: the signal propagates through documented evidence rather than through immediate human interpretation of a system output.

What propagates in P3.2 is documented uncertainty with decision role \(d_{\mathrm{assure}}\) and an explicit evidential scope.Its object of concern, the conditions under which it was estimated, and its limitations must remain visible when the evidence is used in approval, accountability, validation, or operating-control decisions.

The recurring pattern is \emph{governance uptake of uncertainty evidence}: documented uncertainty is converted into evidence that bounds claims or informs operating constraints (e.g., safety cases~\citep{lee2026constructing}). Two uses are visible in the surveyed literature. \emph{Bounded assurance claims} limits the strength of comparative, acceptance, or performance claims when evaluation evidence remains statistically uncertain. This appears in statistically corrected LLM-as-judge reporting, where calibration data and confidence intervals prevent raw evaluator scores from being treated as precise or unbiased measures of system quality~\citep{lee2025correctly}. \emph{Operating-constraint setting} uses drift, inconsistency, or validation evidence to define controls such as model-tier restrictions, deterministic configuration requirements, invariant checks, monitoring, or human review triggers~\citep{khatchadourian2025llm}. Across both uses, uncertainty propagates as governance-usable evidence: it qualifies what can be claimed about the system and the conditions under which its use can be authorised or constrained.

A useful boundary case is uncertainty that is logged or monitored within the technical system without being qualified as evidence for a governance or authorisation decision; such work belongs in P2 rather than P3.2.

\section{Insights and implications}
\label{sec:insights}

Reading across the taxonomy reveals patterns that are not apparent from any single subtype in isolation: recurring failure modes, structural asymmetries, and design tensions that cut across P1, P2, and P3. This section draws out the main cross-cutting insights and their implications for system design, evaluation, and governance.

\subsection{Cross-cutting insights}
\label{sec:cross_cutting_insights}

\paragraph{Propagation is defined by consequential reuse, not mere movement.}
Uncertainty does not meaningfully propagate merely because a score, cue, or estimate appears in multiple places. It propagates when a later consumer takes it up and uses it to shape what happens next. A cue that serves a diagnostic role at one level may trigger runtime control at another, or serve as evidence for governance at a third. What changes across the taxonomy is not only where uncertainty appears, but the role it comes to play and the decisions it enables or forecloses. The uptake context is as important as the signal itself, which is why superficially similar uncertainty signals can support very different downstream interpretations.

\paragraph{The most consequential propagation sites are handoff boundaries, control points, and persistence points.}
Not all sites of uncertainty movement are equally important. \emph{Handoff boundaries} matter because uncertainty must be reformulated for a new component, consumer, or audience, and reformulation can silently alter the object of concern, decision role, or scope of the signal (\hyperref[sec:p23]{P2.3}). \emph{Control points} matter because uncertainty is translated into action, such as selection, escalation, abstention, or routing, and the adequacy of the signal for the control decision it triggers is not always verified (\hyperref[sec:p22]{P2.2}). \emph{Persistence points} matter because transient local uncertainty can be retained in workflow state, memory, or system configuration, where it shapes later steps, later executions, or later users in ways that are no longer traceable to the original estimation context (\hyperref[sec:p24]{P2.4}). These are the sites at which local uncertainty most often acquires system-level consequences, and they are the natural targets for both engineering intervention and evaluation.

\paragraph{Propagation failures are often semantic before they are numeric.}
A recurring pattern across the taxonomy is that the main failure modes are not exhausted by poor estimation or weak calibration. Equally important are failures of meaning. Uncertainty may drift away from the object it originally qualified: the risk identified in meaning-preserving model handoff (\hyperref[sec:p232]{P2.3.2}) is precisely that a stage-local verifier score can be reused downstream as if it were a broader reliability claim~\citep{lee2025correctly}. It may be consumed in a stronger decision role than it supports, a risk illustrated by settings where retrieval coverage diagnostics are treated as sufficient conditions for proceeding rather than as signals that may warrant retry~\citep{soudani2025uncertainty,li2024uncertaintyrag}. Or it may be overgeneralised from a local artefact to a system-wide guarantee, a risk particularly salient at P3 where documented uncertainty enters governance processes without explicit scope conditions, as the mechanisms in P3.2 suggest~\citep{khatchadourian2025llm}. In each case, the uncertainty signal was present; what failed was the preservation of its intended scope and use across propagation. Evaluation that stops at local signal quality will systematically miss these failures.

\paragraph{The research literature is asymmetrically developed across P1, P2, and P3.}
The taxonomy exposes an asymmetry: P1 mechanisms are the most technically developed, with a rich and growing literature on within-request estimation, transformation, and control. P2 mechanisms are moderately covered, particularly routing and re-expression subtypes. P3 mechanisms are substantially thinner, with limited systematic work on how uncertainty shapes human reliance, organisational decision-making, or governance outcomes in deployed LLM systems. This asymmetry likely reflects the greater tractability of model-internal uncertainty estimation compared with the empirical study of socio-technical uptake, where controlled measurement is harder and the relevant outcomes are more distal. The practical consequence is that the propagation mechanisms most directly affecting consequential decisions remain the least understood.

\paragraph{Richer signals at P1 are harder to propagate at P2.}
A tension visible across \hyperref[sec:p23]{P2.3} subtypes is that the more expressive and precise the uncertainty signal produced at P1, the more difficult it becomes to pass across component boundaries without loss. Simple scalar confidence scores propagate easily but carry limited information about scope, coverage, or conditionality. Richer signal forms, such as probability intervals, credal sets, or structured diagnostics, preserve distinctions that scalars collapse~\citep{yang2026verbalizing}, but are harder to encode in boundary-compatible forms that receiving components can act on, as the interface encoding challenges documented in \hyperref[sec:p231]{P2.3.1} and \hyperref[sec:p233]{P2.3.3} illustrate. Increasing local signal richness can impede downstream propagation by creating interface incompatibility that simpler signals avoid. Interface design must therefore be treated as a primary engineering concern alongside estimation quality.

\paragraph{Cross-run adaptation creates feedback loops invisible to within-run evaluation.}
The \hyperref[sec:p24]{P2.4} mechanisms show that uncertainty observed in one run can reshape the artefact pools and decision logic that govern later runs. This creates feedback loops that are structurally invisible to evaluation frameworks that assess system behaviour within a single run or on a fixed held-out set. A system that adapts cautiously under uncertainty may appear well-calibrated in snapshot evaluation while its behaviour shifts in deployment as accumulated uncertainty-conditioned adaptations compound across runs, as the decision-logic adaptation and artefact curation mechanisms in \hyperref[sec:p241]{P2.4.1} and \hyperref[sec:p242]{P2.4.2} illustrate.

\subsection{Implications: from estimating uncertainty to engineering uncertainty}
\label{sec:implications}

A central implication of the taxonomy is that uncertainty in LLM systems should be treated not only as something to estimate, but as something to engineer. The taxonomy shows that uncertainty becomes consequential through the propagation chains by which it is represented, reformulated, reused, retained, disclosed, and constrained for later consumers and decisions. This shifts attention from local uncertainty quality alone to the engineering of the pathways through which uncertainty acquires operational, organisational, and assurance consequences.

\paragraph{Implications for system design.}
The main design implication is to engineer uncertainty for the next consumer, not only for the current producer. At handoff boundaries (\hyperref[sec:p23]{P2.3}), this requires explicit interface semantics that preserve the object of concern and decision role of the signal, so that a stage-local verifier score is not inadvertently reused as a broader reliability claim. At control points (\hyperref[sec:p22]{P2.2}), it requires matching the signal to the decision it triggers: a diagnostic signal is not automatically adequate for a routing or escalation decision, and the adequacy of the match should be specified at the component interface. At persistence points (\hyperref[sec:p24]{P2.4}), it requires disciplined treatment of what uncertainty is written into memory, workflow state, or system configuration, since retained uncertainty shapes future behaviour in ways that are difficult to audit after the fact. Uncertainty should be treated as part of the system interface contract, not as incidental metadata.

\paragraph{Implications for evaluation.}
The main evaluation implication is to assess propagation quality, not only estimate quality. It is not sufficient to ask whether a local signal is plausible, calibrated, or internally useful. Evaluation should also test whether uncertainty remains interpretable after handoff, whether uncertainty-triggered control is proportionate and justified, whether retained uncertainty improves subsequent behaviour without introducing hidden drift, and whether human-facing or assurance-facing disclosures support appropriate reliance rather than overtrust or undertrust. Propagation-aware evaluation follows uncertainty through the handoff boundaries, control points, and persistence mechanisms at which it becomes consequential, rather than stopping at the site where it is first produced. For cross-run mechanisms (\hyperref[sec:p24]{P2.4}), this requires longitudinal evaluation designs that can detect drift introduced by accumulated uncertainty-conditioned adaptation across runs.

\paragraph{Implications for assurance and governance.}
The main assurance implication is that uncertainty should constrain claims, not merely accompany them. When uncertainty is taken up in validation, audit, deployment approval, or governance processes (\hyperref[sec:p32]{P3.2}), it should bound what can be claimed, indicate when additional review is required, and shape the operating conditions under which the system is considered acceptable. This requires assurance artefacts that preserve not only the presence of uncertainty but also its scope, provenance, and the conditions under which it was estimated. The practical target is assurance evidence whose limitations are explicit enough that a governance decision-maker can determine what the evidence does and does not support, rather than treating documented uncertainty as unconditional evidence of system reliability.

\section{Open challenges and future directions}
\label{sec:gaps}

The insights in Section~\ref{sec:insights} show that the main unsolved problems in uncertainty propagation are not reducible to improving local estimation quality. They concern whether uncertainty survives reuse across boundaries with its meaning intact, whether propagation quality can be evaluated rather than merely estimated locally, how preserved uncertainty gets operationalised into system decisions, and how it is taken up in human and institutional processes where its consequences are most direct. We organise the research agenda around these four challenges, adding a fifth on formal foundations which the descriptive scope of the current taxonomy does not address.

\subsection{Preserving semantic fidelity across boundaries}
\label{sec:gaps_semantics}

A first challenge is to preserve semantic fidelity when locally produced uncertainty is reused elsewhere. Across the surveyed literature, uncertainty is operationalised through confidence scores, disagreement measures, entropy-like quantities, validation signals, verifier outcomes, and interface cues, but the associated object of concern and decision role are often left implicit. As a result, signals that are meaningful in one local setting may later be consumed as though their interpretation were unambiguous, even when the downstream consumer faces a different question, acts at a different level of abstraction, or uses the signal for a different purpose.

Three failure modes follow directly. \emph{Semantic drift} occurs when a re-expressed quantity no longer carries the same interpretive content after transformation or handoff. \emph{Object drift} occurs when uncertainty about one object, such as a retrieved passage, intermediate artefact, or local model behaviour, is later read as uncertainty about a different object, such as end-to-end answer quality or broader system adequacy. \emph{Role mismatch} occurs when a quantity introduced for diagnosis or ranking is later reused as a gate, escalation trigger, or assurance input without a clear account of whether it supports that stronger downstream role. These are not merely terminological problems: they determine whether propagated uncertainty remains suited in scope, role, and evidential value to the next decision it is used to support.

Future work should therefore focus less on whether uncertainty is present at a boundary and more on whether its interpretation survives that boundary in a form suitable for downstream use. This calls for uncertainty signal forms that preserve scope, provenance, object, and intended role more explicitly across reuse, rather than leaving these to be inferred later. In some settings this may require signal forms richer than scalar confidence alone, for example interval-, set-, or possibility-based forms that preserve ambiguity, ignorance, or disagreement structure rather than collapsing them prematurely into a point estimate~\citep{yang2026verbalizing}. The aim, however, is not representational richness for its own sake, but propagated uncertainty whose meaning and evidential value remain intelligible after reuse.

\subsection{Evaluating propagation quality}
\label{sec:gaps_evaluation}

A second challenge is that the field evaluates local uncertainty far more often than it evaluates propagation quality. Local calibration, confidence ranking, and final-task performance can all be informative, but they do not by themselves show whether uncertainty remains interpretable and appropriately scoped after re-expression, storage, aggregation, disclosure, or repeated reuse. Different propagation settings expose different failure modes: within-request transformations may compress useful distinctions, cross-component handoffs may weaken provenance and intended use, and external disclosure may alter reliance or oversight behaviour. Evaluation that targets only one local stage is therefore unlikely to reveal whether propagated uncertainty remains fit for downstream use.

This is also an observability problem. Across the literature, it is often difficult to determine where an uncertainty signal was transformed, what part of it was retained, which later component consumed it, and whether its original scope and role remained intact. These difficulties are amplified in advanced LLM systems involving multiple models, tools, agents, persistent memory, and post hoc documentation for review or governance. Without stronger support for tracing, documenting, and inspecting uncertainty across boundaries, many propagation failures remain difficult to detect and harder still to compare systematically.

Future work should treat evaluation and observability as parts of a single propagation-quality agenda. Methodologically, this means moving beyond local estimator quality toward boundary-aware evaluation of whether uncertainty remains interpretable and appropriately scoped after transformation and reuse. Practically, it calls for observability mechanisms and documentation practices that record how uncertainty is re-expressed, what assumptions attach to it, and which downstream consumers rely on it. Real deployments do not stop at one boundary crossing: uncertainty disclosed to users or recorded in assurance artefacts may later feed back into model updates, workflow redesign, or governance constraints, and understanding these cycles requires evaluation methods that can follow uncertainty across connected technical and socio-technical boundaries rather than only at isolated local points.

\subsection{Operationalising propagated uncertainty into system decisions}
\label{sec:gaps_operational}

A third challenge concerns specifically how uncertainty that has survived a boundary crossing gets operationalised into system decisions. This is distinct from Challenge 1, which concerns whether meaning is preserved at the boundary: even when an uncertainty signal arrives at a downstream component with its scope and role intact, the system must still decide how to act on it. Across the \hyperref[sec:cross-component]{P2} literature, uncertainty is carried across workflow steps, re-expressed at component interfaces, exchanged through interaction protocols, and retained in persistent state, yet the logic by which later components consume that uncertainty remains weakly specified.

The central gap is that uncertainty-to-action semantics are implicit in most current systems. The same signal may be used for filtering in one system, routing in another, or abstention in a third, without a clear account of what makes those downstream uses appropriate. The field lacks explicit principles for when propagated uncertainty should support acceptance, deferment, refinement, or refusal, and under what conditions a signal that was adequate for a diagnostic role remains adequate for a stronger control role. A signal that has lost provenance or scope information through boundary crossing should arguably not be promoted to a control role at all, yet current systems rarely enforce such constraints.

Future work should focus on making these action semantics explicit. One direction is comparative system-level study that holds the uncertainty source relatively fixed while varying how that uncertainty is operationally consumed through interface structure, control policy, memory design, or interaction protocol, measuring the downstream consequences of each choice. A second is the development of explicit decision criteria that specify, for a given signal type and propagation context, which downstream actions the signal can and cannot support. A third is workflow design that treats acceptance criteria and execution commitment conditions as first-class parts of the system contract rather than runtime inferences~\citep{lu2026verification}.

\subsection{Beyond the system boundary: human and institutional uptake}
\label{sec:gaps_assurance}

A fourth challenge concerns what happens once uncertainty leaves the technical system and enters socio-technical decision processes. Two downstream regimes are related but distinct. The first is human uptake: how communicated uncertainty affects user reliance, reviewer judgement, intervention, or oversight in specific interaction settings. The second is institutional uptake: how documented uncertainty enters assurance arguments, accountability records, approval processes, operating constraints, or governance decisions.

Our review suggests a clearer and more cumulative evidence base for the first regime than for the second. The human uptake literature documents effects of linguistic and visual uncertainty communication on reliance and task accuracy~\citep{kim2024m,sevastjanova2025layerflow}, but important questions remain: how these effects vary across user expertise, task stakes, and interaction modality; whether disclosed uncertainty supports calibrated reliance or merely shifts overconfidence from system outputs to disclosed signals; and how uncertainty should be communicated when it is multi-source, structured, or conditional rather than scalar. Methodologically, the field needs controlled studies that measure downstream behaviour rather than stated preference, longitudinal designs that track how reliance patterns shift as users gain experience with uncertainty-disclosing systems, and studies that examine oversight and intervention behaviour rather than only immediate reliance judgements.

For institutional uptake, the evidence base is thinner still. Even where uncertainty appears in audit records, validation evidence, or accountability documentation, the literature provides limited guidance on how such material should shape operating envelopes, review thresholds, deployment conditions, or post-deployment obligations. Future work should examine what kinds of claims uncertainty-bearing artefacts can and cannot support, how scope, provenance, and estimation conditions should be preserved in governance-facing documentation, and what institutional review processes are needed to prevent documented uncertainty from being interpreted more strongly than its evidential basis warrants. Progress here will require contributions from both technical communities, which can specify what uncertainty evidence means and what it does not, and governance communities, which can determine how such evidence should be weighted in authorisation and accountability decisions.

A related and currently underexplored dimension concerns how uncertainty propagates through interaction structures in multi-party systems. Early work suggests that communication topology, including who receives an uncertainty signal, how it is retransmitted among technical agents, and how contributions are weighted during aggregation, can amplify, damp, or distort the evidential content of propagated uncertainty~\citep{shen2025understanding,hayashi2025decentralized}. Current evidence is preliminary and has not yet motivated a dedicated taxonomy entry, but this interaction-structural dimension is a natural extension of the multi-agent re-expression mechanisms in P2.3.4 and warrants systematic investigation as multi-agent deployments mature.

\subsection{Formal foundations for uncertainty propagation}
\label{sec:gaps_formal}
A fifth challenge is the absence of formal foundations for uncertainty propagation across heterogeneous system boundaries. The taxonomy is deliberately descriptive, which is appropriate for a survey at the current state of the field, but it leaves open questions that a descriptive taxonomy cannot answer.

Three formal questions are especially pressing. First, what conditions guarantee that an uncertainty signal remains calibrated after aggregation across P1-to-P2 boundaries? The aggregation mechanisms in \hyperref[sec:p121]{P1.2.1} and \hyperref[sec:p21]{P2.1} combine signals in ways whose calibration properties are not generally understood, and standard results from ensemble calibration do not transfer directly to the heterogeneous signal types and aggregation schemes found in compound LLM systems. Second, what properties should hold for uncertainty to compose correctly across model-to-model handoffs? Two desiderata are distinct: calibration preservation, which concerns whether the composed signal remains well-calibrated against outcomes, and scope preservation, which concerns whether the composed signal retains a valid relationship to the object of concern and decision role of its constituents. The meaning-preserving handoff mechanisms in P2.3.2 assume both but provide no formal conditions under which either holds. Third, under what conditions does a signal that is well-formed at production remain well-formed after transformation, re-expression, or retention across the P2 mechanisms surveyed? Here, well-formedness means that the signal retains an interpretable and valid relationship to its object of concern, scope, and intended decision role.

Recent work has begun to provide formal statistical guarantees for uncertainty estimates within specific models of agent behaviour, for example by bounding violation-probability estimates used in runtime control decisions~\citep{wang2026probguard}. However, such guarantees apply within a single formalism over a homogeneous state space. Extending them to heterogeneous cross-component propagation, where signal types, formalisms, and consumers change at each boundary, remains open. Progress may draw on formal frameworks for compositional reasoning under uncertainty, imprecise probability as a foundation for richer signal forms, and belief revision as a basis for cross-run adaptation. The open problem is to adapt these foundations to multi-component, heterogeneous, and governance-facing LLM systems, where uncertainty does not merely update beliefs within a closed model.

\section{Conclusion}
\label{sec:conclusion}

Uncertainty in LLM systems cannot be understood adequately by looking only at isolated estimates attached to single model outputs. In contemporary deployments, uncertainty is produced, transformed, externalised, and reused across model internals, workflow stages, component boundaries, persistent state, and socio-technical processes. The central question is therefore not only whether uncertainty can be estimated locally, but whether its meaning, scope, and decision relevance survive reuse across boundaries.

This survey examined that broader problem as one of \emph{uncertainty propagation}. We introduced a conceptual framing that treats propagated uncertainty as a contextualised signal rather than a score in isolation, and used it to organise the literature across three propagation regions: intra-model propagation within a single model-facing request (P1), system-level propagation across the deployed technical system (P2), and socio-technical propagation beyond the system boundary (P3). Across these regions, uncertainty propagation emerges fundamentally as a boundary problem: as uncertainty crosses boundaries, it may be re-expressed, combined, retained, disclosed, or documented in ways that change both what it is taken to be about and what decisions it is used to support.
The survey also reveals a systematic asymmetry: the mechanisms closest to model internals are the best understood, while those where propagated uncertainty most directly affects consequential human and institutional decisions remain the least studied.

The challenge for the field is not simply to design better local uncertainty estimators, but to ensure that propagated uncertainty remains interpretable, decision-appropriate, and evidentially bounded. This, in turn, requires propagation-aware representations, evaluation methods, observability mechanisms, runtime control designs, and assurance practices that treat uncertainty as a first-class system artefact. As LLM systems become more stateful, tool-using, multi-component, and institutionally embedded, progress will depend less on introducing ever more local signals than on engineering how uncertainty is represented, propagated, evaluated, and governed across the full lifecycle.

\bibliographystyle{unsrt}  
\bibliography{references}

@article{chen2025standalone,
  title={From Standalone LLMs to Integrated Intelligence: A Survey of Compound Al Systems},
  author={Chen, Jiayi and Ye, Junyi and Wang, Guiling},
  journal={arXiv preprint arXiv:2506.04565},
  year={2025}
}

@book{bass2025engineering,
  title={Engineering AI systems: architecture and DevOps essentials},
  author={Bass, Len and Lu, Qinghua and Weber, Ingo and Zhu, Liming},
  year={2025},
  publisher={Addison-Wesley Professional}
}

@article{lewis2020retrieval,
  title={Retrieval-augmented generation for knowledge-intensive nlp tasks},
  author={Lewis, Patrick and Perez, Ethan and Piktus, Aleksandra and Petroni, Fabio and Karpukhin, Vladimir and Goyal, Naman and K{\"u}ttler, Heinrich and Lewis, Mike and Yih, Wen-tau and Rockt{\"a}schel, Tim and others},
  journal={Advances in neural information processing systems},
  volume={33},
  pages={9459--9474},
  year={2020}
}

@inproceedings{li2025review,
  title={A review of prominent paradigms for llm-based agents: Tool use, planning (including rag), and feedback learning},
  author={Li, Xinzhe},
  booktitle={Proceedings of the 31st International Conference on Computational Linguistics},
  pages={9760--9779},
  year={2025}
}

@article{zheng2023judging,
  title={Judging llm-as-a-judge with mt-bench and chatbot arena},
  author={Zheng, Lianmin and Chiang, Wei-Lin and Sheng, Ying and Zhuang, Siyuan and Wu, Zhanghao and Zhuang, Yonghao and Lin, Zi and Li, Zhuohan and Li, Dacheng and Xing, Eric and others},
  journal={Advances in neural information processing systems},
  volume={36},
  pages={46595--46623},
  year={2023}
}

@inproceedings{zhugeagent,
  title={Agent-as-a-Judge: Evaluate Agents with Agents},
  author={Zhuge, Mingchen and Zhao, Changsheng and Ashley, Dylan R and Wang, Wenyi and Khizbullin, Dmitrii and Xiong, Yunyang and Liu, Zechun and Chang, Ernie and Krishnamoorthi, Raghuraman and Tian, Yuandong and others},
  booktitle={Forty-second International Conference on Machine Learning},
year={2025}
}

@inproceedings{yao2022react,
  title={React: Synergizing reasoning and acting in language models},
  author={Yao, Shunyu and Zhao, Jeffrey and Yu, Dian and Du, Nan and Shafran, Izhak and Narasimhan, Karthik R and Cao, Yuan},
  booktitle={The eleventh international conference on learning representations},
  year={2022}
}

@article{shinn2023reflexion,
  title={Reflexion: Language agents with verbal reinforcement learning},
  author={Shinn, Noah and Cassano, Federico and Gopinath, Ashwin and Narasimhan, Karthik and Yao, Shunyu},
  journal={Advances in Neural Information Processing Systems},
  volume={36},
  pages={8634--8652},
  year={2023}
}

@article{zhou2025shielda,
  title={Shielda: Structured handling of exceptions in llm-driven agentic workflows},
  author={Zhou, Jingwen and Chen, Jieshan and Lu, Qinghua and Zhao, Dehai and Zhu, Liming},
  journal={arXiv preprint arXiv:2508.07935},
  year={2025}
}

@article{zhu2025llm,
  title={Where llm agents fail and how they can learn from failures},
  author={Zhu, Kunlun and Liu, Zijia and Li, Bingxuan and Tian, Muxin and Yang, Yingxuan and Zhang, Jiaxun and Han, Pengrui and Xie, Qipeng and Cui, Fuyang and Zhang, Weijia and others},
  journal={arXiv preprint arXiv:2509.25370},
  year={2025}
}

@article{zhang2025agentracer,
  title={AgenTracer: Who Is Inducing Failure in the LLM Agentic Systems?},
  author={Zhang, Guibin and Wang, Junhao and Chen, Junjie and Zhou, Wangchunshu and Wang, Kun and Yan, Shuicheng},
  journal={arXiv preprint arXiv:2509.03312},
  year={2025}
}

@article{kalai2025language,
  title={Why language models hallucinate},
  author={Kalai, Adam Tauman and Nachum, Ofir and Vempala, Santosh S and Zhang, Edwin},
  journal={arXiv preprint arXiv:2509.04664},
  year={2025}
}

@inproceedings{kalai2024calibrated,
  title={Calibrated language models must hallucinate},
  author={Kalai, Adam Tauman and Vempala, Santosh S},
  booktitle={Proceedings of the 56th Annual ACM Symposium on Theory of Computing},
  pages={160--171},
  year={2024}
}

@article{chhikara2025mind,
  title={Mind the confidence gap: Overconfidence, calibration, and distractor effects in large language models},
  author={Chhikara, Prateek},
  journal={Transactions on Machine Learning Research},
  year={2025}
}

@inproceedings{xiongcan,
  title={Can LLMs Express Their Uncertainty? An Empirical Evaluation of Confidence Elicitation in LLMs},
  author={Xiong, Miao and Hu, Zhiyuan and Lu, Xinyang and LI, YIFEI and Fu, Jie and He, Junxian and Hooi, Bryan},
  booktitle={The Twelfth International Conference on Learning Representations},
  year={2024}
}

@inproceedings{zhao2025uncertainty,
  title={Uncertainty propagation on llm agent},
  author={Zhao, Qiwei and Li, Dong and Liu, Yanchi and Cheng, Wei and Sun, Yiyou and Oishi, Mika and Osaki, Takao and Matsuda, Katsushi and Yao, Huaxiu and Zhao, Chen and others},
  booktitle={Proceedings of the 63rd Annual Meeting of the Association for Computational Linguistics (Volume 1: Long Papers)},
  pages={6064--6073},
  year={2025}
}

@article{shorinwa2025survey,
  title={A survey on uncertainty quantification of large language models: Taxonomy, open research challenges, and future directions},
  author={Shorinwa, Ola and Mei, Zhiting and Lidard, Justin and Ren, Allen Z and Majumdar, Anirudha},
  journal={ACM Computing Surveys},
  year={2025},
  publisher={ACM New York, NY}
}

@article{fu2025deep,
  title={Deep think with confidence},
  author={Fu, Yichao and Wang, Xuewei and Tian, Yuandong and Zhao, Jiawei},
  journal={arXiv preprint arXiv:2508.15260},
  year={2025}
}

@inproceedings{liu2025uncertainty,
  title={Uncertainty quantification and confidence calibration in large language models: A survey},
  author={Liu, Xiaoou and Chen, Tiejin and Da, Longchao and Chen, Chacha and Lin, Zhen and Wei, Hua},
  booktitle={Proceedings of the 31st ACM SIGKDD Conference on Knowledge Discovery and Data Mining V. 2},
  pages={6107--6117},
  year={2025}
}

@article{zhu2025uncertainty,
  title={Uncertainty-guided chain-of-thought for code generation with llms},
  author={Zhu, Yuqi and Li, Ge and Jiang, Xue and Li, Jia and Mei, Hong and Jin, Zhi and Dong, Yihong},
  journal={arXiv preprint arXiv:2503.15341},
  year={2025}
}

@inproceedings{stoissertowards,
  title={Towards Agents That Know When They Don't Know: Uncertainty as a Control Signal for Structured Reasoning},
  author={Stoisser, Josefa Lia and Martell, Marc Boubnovski and Phillips, Lawrence and Mazzoni, Gianluca and Harder, Lea M{\o}rch and Torr, Philip and Ferkinghoff-Borg, Jesper and M{\"a}rtens, Kaspar and Fauqueur, Julien},
  booktitle={Workshop on Scaling Environments for Agents},
  year={2025}
}

@inproceedings{dengplanu,
  title={PlanU: Large Language Model Reasoning through Planning under Uncertainty},
  author={Deng, Ziwei and Deng, Mian and Liang, Chenjing and Gao, Zeming and Ma, Chennan and Lin, Chenxing and Zhang, Haipeng and Mei, Songzhu and Shen, Siqi and Wang, Cheng},
  booktitle={The Thirty-ninth Annual Conference on Neural Information Processing Systems},
  year={2025}
}

@article{hu2024uncertainty,
  title={Uncertainty of thoughts: Uncertainty-aware planning enhances information seeking in llms},
  author={Hu, Zhiyuan and Liu, Chumin and Feng, Xidong and Zhao, Yilun and Ng, See-Kiong and Luu, Anh Tuan and He, Junxian and Koh, Pang Wei W and Hooi, Bryan},
  journal={Advances in Neural Information Processing Systems},
  volume={37},
  pages={24181--24215},
  year={2024}
}

@article{suri2025structured,
  title={Structured Uncertainty guided Clarification for LLM Agents},
  author={Suri, Manan and Mathur, Puneet and Lipka, Nedim and Dernoncourt, Franck and Rossi, Ryan A and Manocha, Dinesh},
  journal={arXiv preprint arXiv:2511.08798},
  year={2025}
}

@article{xu2025llm,
  title={LLM-Based Agents for Tool Learning: A Survey},
  author={Xu, Weikai and Huang, Chengrui and Gao, Shen and Shang, Shuo},
  journal={Data Science and Engineering},
  pages={1--31},
  year={2025},
  publisher={Springer}
}

@article{soudani2025uncertainty,
  title={Why Uncertainty Estimation Methods Fall Short in RAG: An Axiomatic Analysis},
  author={Soudani, Heydar and Kanoulas, Evangelos and Hasibi, Faegheh},
  journal={arXiv preprint arXiv:2505.07459},
  year={2025}
}

@article{li2024uncertaintyrag,
  title={Uncertaintyrag: Span-level uncertainty enhanced long-context modeling for retrieval-augmented generation},
  author={Li, Zixuan and Xiong, Jing and Ye, Fanghua and Zheng, Chuanyang and Wu, Xun and Lu, Jianqiao and Wan, Zhongwei and Liang, Xiaodan and Li, Chengming and Sun, Zhenan and others},
  journal={arXiv preprint arXiv:2410.02719},
  year={2024}
}

@article{jiang2026sok,
  title={SoK: Agentic Skills--Beyond Tool Use in LLM Agents},
  author={Jiang, Yanna and Li, Delong and Deng, Haiyu and Ma, Baihe and Wang, Xu and Wang, Qin and Yu, Guangsheng},
  journal={arXiv preprint arXiv:2602.20867},
  year={2026}
}

@misc{zhang_cot-uq_2025,
    title = {{CoT}-{UQ}: {Improving} {Response}-wise {Uncertainty} {Quantification} in {LLMs} with {Chain}-of-{Thought}},
    shorttitle = {{CoT}-{UQ}},
    url = {http://arxiv.org/abs/2502.17214},
    doi = {10.48550/arXiv.2502.17214},
    language = {en-US},
    urldate = {2025-11-21},
    publisher = {arXiv},
    author = {Zhang, Boxuan and Zhang, Ruqi},
    month = jun,
    year = {2025},
}

@article{huang2024survey,
  title={A survey of uncertainty estimation in llms: Theory meets practice},
  author={Huang, Hsiu-Yuan and Yang, Yutong and Zhang, Zhaoxi and Lee, Sanwoo and Wu, Yunfang},
  journal={arXiv preprint arXiv:2410.15326},
  year={2024}
}

@article{xia2025survey,
  title={A survey of uncertainty estimation methods on large language models},
  author={Xia, Zhiqiu and Xu, Jinxuan and Zhang, Yuqian and Liu, Hang},
  journal={arXiv preprint arXiv:2503.00172},
  year={2025}
}

@article{he2025survey,
  title={Survey of Uncertainty Estimation in Large Language Models-Sources, Methods, Applications, and Challenge},
  author={He, Jianfeng and Yu, Linlin and Li, Changbin and Yang, Runing and Chen, Fanglan and Li, Kangshuo and Zhang, Min and Lei, Shuo and Zhang, Xuchao and Beigi, Mohammad and others},
  year={2025}
}

@article{kang2025uncertainty,
  title={Uncertainty Quantification for Hallucination Detection in Large Language Models: Foundations, Methodology, and Future Directions},
  author={Kang, Sungmin and Bakman, Yavuz Faruk and Yaldiz, Duygu Nur and Buyukates, Baturalp and Avestimehr, Salman},
  journal={arXiv preprint arXiv:2510.12040},
  year={2025}
}

@article{ye2023cognitive,
  title={Cognitive mirage: A review of hallucinations in large language models},
  author={Ye, Hongbin and Liu, Tong and Zhang, Aijia and Hua, Wei and Jia, Weiqiang},
  journal={arXiv preprint arXiv:2309.06794},
  year={2023}
}

@article{zhang2025siren,
    title = {{Siren}’s {Song} in the {AI} {Ocean}: {A} {Survey} on {Hallucination} in {Large} {Language} {Models}},
    url = {https://direct.mit.edu/coli/article/doi/10.1162/coli.a.16/131631},
    urldate = {2026-01-15},
    journal = {Computational Linguistics},
    author = {Zhang, Yue and Li, Yafu and Cui, Leyang and Cai, Deng and Liu, Lemao and Fu, Tingchen and Huang, Xinting and Zhao, Enbo and Zhang, Yu and Chen, Yulong},
    year = {2025},
    pages = {1--46},
}

@article{huang2025hallucination,
author = {Huang, Lei and Yu, Weijiang and Ma, Weitao and Zhong, Weihong and Feng, Zhangyin and Wang, Haotian and Chen, Qianglong and Peng, Weihua and Feng, Xiaocheng and Qin, Bing and Liu, Ting},
title = {A Survey on Hallucination in Large Language Models: Principles, Taxonomy, Challenges, and Open Questions},
year = {2025},
issue_date = {March 2025},
publisher = {Association for Computing Machinery},
address = {New York, NY, USA},
volume = {43},
number = {2},
issn = {1046-8188},
url = {https://doi.org/10.1145/3703155},
doi = {10.1145/3703155},
journal = {ACM Trans. Inf. Syst.},
month = jan,
articleno = {42},
numpages = {55},
}

@inproceedings{pan2025towards,
  title={Towards reliable large language models: A survey on hallucination detection},
  author={Pan, Yao and Kong, Linggang and Wu, Jiaju and Yang, Yonghui and Zuo, Hongfu and Xiu, Ze and Wang, Xiaodong},
  booktitle={International Conference on Intelligent Computing},
  pages={438--451},
  year={2025},
  organization={Springer}
}

@article{tonmoy2024survey,
      title={A Comprehensive Survey of Hallucination Mitigation Techniques in Large Language Models}, 
      author={S. M Towhidul Islam Tonmoy and S M Mehedi Zaman and Vinija Jain and Anku Rani and Vipula Rawte and Aman Chadha and Amitava Das},
      year={2024},
  journal={arXiv preprint arXiv:2401.01313}
}

@article{gao2024ragllm,
      title={Retrieval-Augmented Generation for Large Language Models: A Survey}, 
      author={Yunfan Gao and Yun Xiong and Xinyu Gao and Kangxiang Jia and Jinliu Pan and Yuxi Bi and Yi Dai and Jiawei Sun and Meng Wang and Haofen Wang},
      year={2024},
  journal={arXiv preprint arXiv:2312.10997} 
}

@article{wang_survey_2024,
    title = {A survey on large language model based autonomous agents},
    volume = {18},
    issn = {2095-2228, 2095-2236},
    url = {https://link.springer.com/10.1007/s11704-024-40231-1},
    doi = {10.1007/s11704-024-40231-1},
    language = {en},
    number = {6},
    urldate = {2024-06-06},
    journal = {Frontiers of Computer Science},
    author = {Wang, Lei and Ma, Chen and Feng, Xueyang and Zhang, Zeyu and Yang, Hao and Zhang, Jingsen and Chen, Zhiyuan and Tang, Jiakai and Chen, Xu and Lin, Yankai and Zhao, Wayne Xin and Wei, Zhewei and Wen, Jirong},
    month = dec,
    year = {2024},
    pages = {186345},
}

@inproceedings{guo2024multiagent,
author = {Guo, Taicheng and Chen, Xiuying and Wang, Yaqi and Chang, Ruidi and Pei, Shichao and Chawla, Nitesh V. and Wiest, Olaf and Zhang, Xiangliang},
title = {Large language model based multi-agents: a survey of progress and challenges},
year = {2024},
isbn = {978-1-956792-04-1},
url = {https://doi.org/10.24963/ijcai.2024/890},
doi = {10.24963/ijcai.2024/890},
booktitle = {Proceedings of the Thirty-Third International Joint Conference on Artificial Intelligence},
articleno = {890},
numpages = {10},
location = {Jeju, Korea},
series = {IJCAI '24}
}

@inproceedings{fan2024survey,
  title={A survey on rag meeting llms: Towards retrieval-augmented large language models},
  author={Fan, Wenqi and Ding, Yujuan and Ning, Liangbo and Wang, Shijie and Li, Hengyun and Yin, Dawei and Chua, Tat-Seng and Li, Qing},
  booktitle={Proceedings of the 30th ACM SIGKDD conference on knowledge discovery and data mining},
  pages={6491--6501},
  year={2024}
}

@article{huang2024understanding,
  title={Understanding the planning of LLM agents: A survey},
  author={Huang, Xu and Liu, Weiwen and Chen, Xiaolong and Wang, Xingmei and Wang, Hao and Lian, Defu and Wang, Yasheng and Tang, Ruiming and Chen, Enhong},
  journal={arXiv preprint arXiv:2402.02716},
  year={2024}
}

@article{xi2025rise,
  title={The rise and potential of large language model based agents: A survey},
  author={Xi, Zhiheng and Chen, Wenxiang and Guo, Xin and He, Wei and Ding, Yiwen and Hong, Boyang and Zhang, Ming and Wang, Junzhe and Jin, Senjie and Zhou, Enyu and others},
  journal={Science China Information Sciences},
  volume={68},
  number={2},
  pages={121101},
  year={2025},
  publisher={Springer}
}

@article{vazhentsev2025uncertainty,
  title={Uncertainty-Aware Attention Heads: Efficient Unsupervised Uncertainty Quantification for LLMs},
  author={Vazhentsev, Artem and Rvanova, Lyudmila and Kuzmin, Gleb and Fadeeva, Ekaterina and Lazichny, Ivan and Panchenko, Alexander and Panov, Maxim and Baldwin, Timothy and Sachan, Mrinmaya and Nakov, Preslav and others},
  journal={arXiv preprint arXiv:2505.20045},
  year={2025}
}

@article{moslonka2025learned,
  title={Learned hallucination detection in black-box llms using token-level entropy production rate},
  author={Moslonka, Charles and Randrianarivo, Hicham and Garnier, Arthur and Malherbe, Emmanuel},
  journal={arXiv preprint arXiv:2509.04492},
  year={2025}
}

@article{lymperopoulos2025tools,
  title={Tools in the Loop: Quantifying Uncertainty of LLM Question Answering Systems That Use Tools},
  author={Lymperopoulos, Panagiotis and Sarathy, Vasanth},
  journal={arXiv preprint arXiv:2505.16113},
  year={2025}
}

@article{huang2025reppl,
  title={RePPL: Recalibrating Perplexity by Uncertainty in Semantic Propagation and Language Generation for Explainable QA Hallucination Detection},
  author={Huang, Yiming and Zhang, Junyan and Wang, Zihao and Bie, Biquan and Qiu, Yunzhong and Fung, Yi R and He, Xinlei},
  journal={arXiv preprint arXiv:2505.15386},
  year={2025}
}

@inproceedings{yuan2025kg,
  title={KG-UQ: Knowledge Graph-Based Uncertainty Quantification for Long Text in Large Language Models},
  author={Yuan, Yingqing and Tao, Linwei and Lu, Haohui and Khushi, Matloob and Razzak, Imran and Dras, Mark and Yang, Jian and Naseem, Usman},
  booktitle={Companion Proceedings of the ACM on Web Conference 2025},
  pages={2071--2077},
  year={2025}
}

@article{yoffe2024debunc,
  title={DebUnc: Improving Large Language Model Agent Communication With Uncertainty Metrics},
  author={Yoffe, Luke and Amayuelas, Alfonso and Wang, William Yang},
  journal={arXiv preprint arXiv:2407.06426},
  year={2024}
}

@article{zhou2025can,
  title={Can LLMs Detect Their Confabulations? Estimating Reliability in Uncertainty-Aware Language Models},
  author={Zhou, Tianyi and Medina, Johanne and Chawla, Sanjay},
  journal={arXiv preprint arXiv:2508.08139},
  year={2025}
}

@inproceedings{lee2025uncertainty,
  title={Uncertainty-Aware Contrastive Decoding},
  author={Lee, Hakyung and Park, Subeen and Kim, Joowang and Lim, Sungjun and Song, Kyungwoo},
  booktitle={Findings of the Association for Computational Linguistics: ACL 2025},
  pages={26376--26391},
  year={2025}
}

@article{arbuzov2025beyond,
  title={Beyond Exponential Decay: Rethinking Error Accumulation in Large Language Models},
  author={Arbuzov, Mikhail L and Shvets, Alexey A and Beir, Sisong},
  journal={arXiv preprint arXiv:2505.24187},
  year={2025}
}

@inproceedings{sevastjanova2025layerflow,
  title={LayerFlow: Layer-wise Exploration of LLM Embeddings using Uncertainty-aware Interlinked Projections},
  author={Sevastjanova, Rita and Gerling, Robin and Spinner, Thilo and El-Assady, Mennatallah},
  booktitle={Computer Graphics Forum},
  pages={e70123},
  year={2025},
  organization={Wiley Online Library}
}

@inproceedings{tang2025analysis,
  title={Analysis of Image-and-Text Uncertainty Propagation in Multimodal Large Language Models with Cardiac MR-Based Applications},
  author={Tang, Yucheng and Fu, Yunguan and Yi, Weixi and Wang, Yipei and Alexander, Daniel C and Davies, Rhodri and Hu, Yipeng},
  booktitle={International Conference on Medical Image Computing and Computer-Assisted Intervention},
  pages={36--45},
  year={2025},
  organization={Springer}
}

@article{ciefadaptive,
  title={Adaptive Uncertainty-Aware Reinforcement Learning from Human Feedback},
  author={Cief, Matej and Tonolini, Francesco and Aletras, Nikolaos and Kazai, Gabriella},
  year={2024}
}

@article{khatchadourian2025llm,
  title={LLM Output Drift: Cross-Provider Validation \& Mitigation for Financial Workflows},
  author={Khatchadourian, Raffi and Franco, Rolando},
  journal={arXiv preprint arXiv:2511.07585},
  year={2025}
}

@inproceedings{chen2025enhancing,
  title={Enhancing uncertainty modeling with semantic graph for hallucination detection},
  author={Chen, Kedi and Chen, Qin and Zhou, Jie and Tao, Xinqi and Ding, Bowen and Xie, Jingwen and Xie, Mingchen and Li, Peilong and Feng, Zheng},
  booktitle={Proceedings of the AAAI Conference on Artificial Intelligence},
  volume={39},
  number={22},
  pages={23586--23594},
  year={2025}
}

@article{lee2025correctly,
  title={How to correctly report llm-as-a-judge evaluations},
  author={Lee, Chungpa and Zeng, Thomas and Jeong, Jongwon and Sohn, Jy-yong and Lee, Kangwook},
  journal={arXiv preprint arXiv:2511.21140},
  year={2025}
}

@inproceedings{kim2024m,
  title={" I'm Not Sure, But...": Examining the Impact of Large Language Models' Uncertainty Expression on User Reliance and Trust},
  author={Kim, Sunnie SY and Liao, Q Vera and Vorvoreanu, Mihaela and Ballard, Stephanie and Vaughan, Jennifer Wortman},
  booktitle={Proceedings of the 2024 ACM conference on fairness, accountability, and transparency},
  pages={822--835},
  year={2024}
}

@article{zellinger2025rational,
  title={Rational tuning of llm cascades via probabilistic modeling},
  author={Zellinger, Michael J and Thomson, Matt},
  journal={arXiv preprint arXiv:2501.09345},
  year={2025}
}

@article{duan2025uprop,
  title={UProp: Investigating the Uncertainty Propagation of LLMs in Multi-Step Agentic Decision-Making},
  author={Duan, Jinhao and Diffenderfer, James and Madireddy, Sandeep and Chen, Tianlong and Kailkhura, Bhavya and Xu, Kaidi},
  journal={arXiv preprint arXiv:2506.17419},
  year={2025}
}

@article{abbasli2025comparing,
  title={Comparing Uncertainty Measurement and Mitigation Methods for Large Language Models: A Systematic Review},
  author={Abbasli, Toghrul and Toyoda, Kentaroh and Wang, Yuan and Witt, Leon and Ali, Muhammad Asif and Miao, Yukai and Li, Dan and Wei, Qingsong},
  journal={arXiv preprint arXiv:2504.18346},
  year={2025}
}

@inproceedings{vazhentsev2025unconditional,
  title={Unconditional truthfulness: Learning unconditional uncertainty of large language models},
  author={Vazhentsev, Artem and Fadeeva, Ekaterina and Xing, Rui and Kuzmin, Gleb and Lazichny, Ivan and Panchenko, Alexander and Nakov, Preslav and Baldwin, Timothy and Panov, Maxim and Shelmanov, Artem},
  booktitle={Proceedings of the 2025 Conference on Empirical Methods in Natural Language Processing},
  pages={35661--35682},
  year={2025}
}

@article{zur2025language,
  title={Are language models aware of the road not taken? Token-level uncertainty and hidden state dynamics},
  author={Zur, Amir and Geiger, Atticus and Lubana, Ekdeep Singh and Bigelow, Eric},
  journal={arXiv preprint arXiv:2511.04527},
  year={2025}
}

@article{tan2025bottom,
  title={Bottom-up policy optimization: Your language model policy secretly contains internal policies},
  author={Tan, Yuqiao and Wang, Minzheng and He, Shizhu and Liao, Huanxuan and Zhao, Chengfeng and Lu, Qiunan and Liang, Tian and Zhao, Jun and Liu, Kang},
  journal={arXiv preprint arXiv:2512.19673},
  year={2025}
}

@article{budzinskiy2025numerical,
  title={Numerical Error Analysis of Large Language Models},
  author={Budzinskiy, Stanislav and Fang, Wenyi and Zeng, Longbin and Petersen, Philipp},
  journal={arXiv preprint arXiv:2503.10251},
  year={2025}
}

@inproceedings{gao2025flue,
  title={FLUE: Streamlined Uncertainty Estimation for Large Language Models},
  author={Gao, Shiqi and Gong, Tianxiang and Lin, Zijie and Xu, Runhua and Zhou, Haoyi and Li, Jianxin},
  booktitle={Proceedings of the AAAI Conference on Artificial Intelligence},
  volume={39},
  number={16},
  pages={16745--16753},
  year={2025}
}

@article{yeh2025halluentity,
  title={Halluentity: Benchmarking and understanding entity-level hallucination detection},
  author={Yeh, Min-Hsuan and Kamachee, Max and Park, Seongheon and Li, Yixuan},
  journal={Transactions on Machine Learning Research},
  year={2025}
}

@article{ghasemabadi2025can,
  title={Can LLMs Predict Their Own Failures? Self-Awareness via Internal Circuits},
  author={Ghasemabadi, Amirhosein and Niu, Di},
  journal={arXiv preprint arXiv:2512.20578},
  year={2025}
}

@inproceedings{dakhmouche2025can,
  title={Can Linear Probes Measure LLM Uncertainty?},
  author={Dakhmouche, Ramzi and Letellier, Adrien and Gorji, Hossein},
  booktitle={NeurIPS 2025 Workshop MLxOR: Mathematical Foundations and Operational Integration of Machine Learning for Uncertainty-Aware Decision-Making},
year={2025}
}

@article{manvi2025zero,
  title={Zero-Overhead Introspection for Adaptive Test-Time Compute},
  author={Manvi, Rohin and Hong, Joey and Seyde, Tim and Labonne, Maxime and Lechner, Mathias and Levine, Sergey},
  journal={arXiv preprint arXiv:2512.01457},
  year={2025}
}

@article{lou2024uncertainty,
  title={Uncertainty-aware reward model: Teaching reward models to know what is unknown},
  author={Lou, Xingzhou and Yan, Dong and Shen, Wei and Yan, Yuzi and Xie, Jian and Zhang, Junge},
  journal={arXiv preprint arXiv:2410.00847},
  year={2024}
}

@inproceedings{zhangtokur,
  title={TokUR: Token-Level Uncertainty Estimation for Large Language Model Reasoning},
  author={Zhang, Tunyu and Shi, Haizhou and Wang, Yibin and Wang, Hengyi and He, Xiaoxiao and Li, Zhuowei and Chen, Haoxian and Han, Ligong and Xu, Kai and Zhang, Huan and others},
  booktitle={First Workshop on Foundations of Reasoning in Language Models},
  year={2025}
}

@article{schuster2022confident,
  title={Confident adaptive language modeling},
  author={Schuster, Tal and Fisch, Adam and Gupta, Jai and Dehghani, Mostafa and Bahri, Dara and Tran, Vinh and Tay, Yi and Metzler, Donald},
  journal={Advances in Neural Information Processing Systems},
  volume={35},
  pages={17456--17472},
  year={2022}
}

@article{yarie2024mitigating,
  title={Mitigating token-level uncertainty in retrieval-augmented large language models},
  author={Yarie, Liz and Soriano, Dominic and Kaczmarek, Leonard and Wilkinson, Benjamin and Vasquez, Eduardo},
  journal={Authorea Preprints},
  publisher={Authorea},
  year={2024}
}

@article{zhang2026agentic,
  title={Agentic Confidence Calibration},
  author={Zhang, Jiaxin and Xiong, Caiming and Wu, Chien-Sheng},
  journal={arXiv preprint arXiv:2601.15778},
  year={2026}
}

@inproceedings{razghandi2025cer,
  author       = {Ali Razghandi and
                  Seyed Mohammad Hadi Hosseini and
                  Mahdieh Soleymani Baghshah},
  title        = {{CER:} Confidence Enhanced Reasoning in LLMs},
  booktitle    = {Proceedings of the 63rd Annual Meeting of the Association for Computational
                  Linguistics (Volume 1: Long Papers), {ACL} 2025, Vienna, Austria,
                  July 27 - August 1, 2025},
  pages        = {7918--7938},
  publisher    = {Association for Computational Linguistics},
  year         = {2025}
}

@article{phillips2025geometric,
  title={Geometric uncertainty for detecting and correcting hallucinations in llms},
  author={Phillips, Edward and Wu, Sean and Molaei, Soheila and Belgrave, Danielle and Thakur, Anshul and Clifton, David},
  journal={arXiv preprint arXiv:2509.13813},
  year={2025}
}

@inproceedings{yin2024reasoning,
  title={Reasoning in flux: Enhancing large language models reasoning through uncertainty-aware adaptive guidance},
  author={Yin, Zhangyue and Sun, Qiushi and Guo, Qipeng and Zeng, Zhiyuan and Li, Xiaonan and Dai, Junqi and Cheng, Qinyuan and Huang, Xuan-Jing and Qiu, Xipeng},
  booktitle={Proceedings of the 62nd Annual Meeting of the Association for Computational Linguistics (Volume 1: Long Papers)},
  pages={2401--2416},
  year={2024}
}

@article{yu2025uncertainty,
  title={Robust Search with Uncertainty-Aware Value Models for Language Model Reasoning},
  author={Yu, Fei and Li, Yingru and Wang, Benyou},
  journal={arXiv preprint arXiv:2502.11155},
  year={2025}
}

@inproceedings{oh2025uncertainty,
  title={Uncertainty-aware hybrid inference with on-device small and remote large language models},
  author={Oh, Seungeun and Kim, Jinhyuk and Park, Jihong and Ko, Seung-Woo and Quek, Tony QS and Kim, Seong-Lyun},
  booktitle={2025 IEEE International Conference on Machine Learning for Communication and Networking (ICMLCN)},
  pages={1--7},
  year={2025},
  organization={IEEE}
}

@inproceedings{renrobots,
  title={Robots That Ask For Help: Uncertainty Alignment for Large Language Model Planners},
  author={Ren, Allen Z and Dixit, Anushri and Bodrova, Alexandra and Singh, Sumeet and Tu, Stephen and Brown, Noah and Xu, Peng and Takayama, Leila and Xia, Fei and Varley, Jake and others},
  booktitle={7th Annual Conference on Robot Learning},
  year={2023}
}

@article{choudhury2025bed,
  title={Bed-llm: Intelligent information gathering with llms and bayesian experimental design},
  author={Choudhury, Deepro and Williamson, Sinead and Goli{\'n}ski, Adam and Miao, Ning and Smith, Freddie Bickford and Kirchhof, Michael and Zhang, Yizhe and Rainforth, Tom},
  journal={arXiv preprint arXiv:2508.21184},
  year={2025}
}

@inproceedings{nafar2025reasoning,
  title={Reasoning over uncertain text by generative large language models},
  author={Nafar, Aliakbar and Venable, Kristen Brent and Kordjamshidi, Parisa},
  booktitle={Proceedings of the AAAI Conference on Artificial Intelligence},
  volume={39},
  number={23},
  pages={24911--24920},
  year={2025}
}

@inproceedings{hayashi2025decentralized,
  title={Decentralized Belief Propagation in LLM Agents: A Brain-Inspired Approach to AI Safety Analysis},
  author={Hayashi, Yusuke},
  booktitle={International Conference on Neural Information Processing},
  pages={537--550},
  year={2025},
  organization={Springer}
}

@article{shen2025understanding,
  title={Understanding the Information Propagation Effects of Communication Topologies in LLM-based Multi-Agent Systems},
  author={Shen, Xu and Liu, Yixin and Dai, Yiwei and Wang, Yili and Miao, Rui and Tan, Yue and Pan, Shirui and Wang, Xin},
  journal={arXiv preprint arXiv:2505.23352},
  year={2025}
}

@inproceedings{yu2023cold,
  title={Cold-start data selection for better few-shot language model fine-tuning: A prompt-based uncertainty propagation approach},
  author={Yu, Yue and Zhang, Rongzhi and Xu, Ran and Zhang, Jieyu and Shen, Jiaming and Zhang, Chao},
  booktitle={Proceedings of the 61st annual meeting of the association for computational linguistics (volume 1: long papers)},
  pages={2499--2521},
  year={2023}
}

@inproceedings{liang2024actively,
  title={Actively learn from llms with uncertainty propagation for generalized category discovery},
  author={Liang, Jinggui and Liao, Lizi and Fei, Hao and Li, Bobo and Jiang, Jing},
  booktitle={Proceedings of the 2024 Conference of the North American Chapter of the Association for Computational Linguistics: Human Language Technologies (Volume 1: Long Papers)},
  pages={7838--7851},
  year={2024}
}

@article{kiyani2026trust,
  title={When to Trust the Cheap Check: Weak and Strong Verification for Reasoning},
  author={Kiyani, Shayan and Noorani, Sima and Pappas, George and Hassani, Hamed},
  journal={arXiv preprint arXiv:2602.17633},
  year={2026}
}

@article{qiu2024llm,
  title={LLM-based agentic systems in medicine and healthcare},
  author={Qiu, Jianing and Lam, Kyle and Li, Guohao and Acharya, Amish and Wong, Tien Yin and Darzi, Ara and Yuan, Wu and Topol, Eric J},
  journal={Nature Machine Intelligence},
  volume={6},
  number={12},
  pages={1418--1420},
  year={2024},
  publisher={Nature Publishing Group UK London}
}

@article{yang2026verbalizing,
  title={Verbalizing LLM’s Higher-order Uncertainty via Imprecise Probabilities},
  author={Yang, Anita and Muandet, Krikamol and Caprio, Michele and Chau, Siu Lun and Adachi, Masaki},
  year={2026},
}

@article{lu2026verification,
  title={Verification-Driven AI Engineering: Workflows and Reference Architecture},
  author={Lu, Qinghua and Zhu, Liming and Ma, Suyu and Power, Helen and Speight, Robert},
  year={2026}
}

@article{lee2026constructing,
  title={Constructing Safety Cases for AI Systems: A Reusable Template Framework},
  author={Lee, Sung Une and Zhu, Liming and Shamsujjoha, Md and Dong, Liming and Lu, Qinghua and Chen, Jieshan},
  journal={arXiv preprint arXiv:2601.22773},
  year={2026}
}

@misc{wang2026probguard,
      title={ProbGuard: Probabilistic Runtime Monitoring for LLM Agent Safety}, 
      author={Haoyu Wang and Christopher M. Poskitt and Jiali Wei and Jun Sun},
      year={2026},
      eprint={2508.00500},
      archivePrefix={arXiv},
      primaryClass={cs.AI},
      url={https://arxiv.org/abs/2508.00500}, 
}

\end{document}